\documentclass{article}%
\usepackage{amsmath}
\usepackage{amsfonts}
\usepackage{amssymb}
\usepackage{graphicx}%
\providecommand{\U}[1]{\protect\rule{.1in}{.1in}}
\newtheorem{theorem}{Theorem}

\newtheorem{lemma}[theorem]{Lemma}

\begin{document}

\title{Modular localization and the foundational origin of integrability \\\ \ {\small dedicated to Raymond Stora on the occasion of his 80th birthday}}
\author{Bert Schroer\\present address: CBPF, Rua Dr. Xavier Sigaud 150, \\22290-180 Rio de Janeiro, Brazil\\email schroer@cbpf.br\\permanent address: Institut f\"{u}r Theoretische Physik\\FU-Berlin, Arnimallee 14, 14195 Berlin, Germany}
\date{November, \ 2012}
\maketitle

\begin{abstract}
The main aim of this work is to relate integrability in QFT with a complete
particle interpretation directly to the principle of causal localization,
circumventing the standard method of finding sufficiently many conservation
laws. Its precise conceptual-mathematical formulation as "modular
localization" within the setting of local operator algebras also suggests
novel ways of looking at general (non-integrable) QFTs which are not based on
quantizing classical field theories.

Conformal QFT, which is known to admit no particle interpretation, suggest the
presence of a "partial" integrability, referred to as "conformal
integrability". This manifest itself in a "braid-permutation" group structure
which contains in particular informations about the anomalous dimensional
spectrum. For chiral conformal models this reduces to the braid group as it is
represented in Hecke- or Birman-Wenzl- algebras associated to chiral models.

Another application of modular localization mentioned in this work is an
alternative to the BRST formulation of gauge theories in terms of stringlike
vectorpotentials within a Hilbert space setting.

\end{abstract}

\section{Integrability in classical and quantum theory}

The notion of integrability arose in celestial mechanics and referred to
systems for which the equation of motion can be solved in closed analytic form
without the necessity to resort to controlled approximations (perturbation
theory). The prototype model was the Kepler two-body system, whereas more than
two celestial bodies lead to non-integrable situations which can only be
approximated (in principle with unlimited numerical accuracy). For the
non-integrable case the terminology does not just mean that no analytic
solution was found, but rather points to the existence of a proof that such a
solution does not exist.

As the mathematical sophistication evolved, physicists and mathematicians
developed model-independent criteria for integrability. A modern definition
which is sufficiently general to cover classical mechanics is in terms of a
\textit{complete set of conservation laws in involution} \cite{Arn}.

This definition was extended from mechanics to include \textit{classical field
theory} where, according to Noether's theorem, a symmetry in the Lagrangian
setting leads to a conserved current and integrability means that there exists
an infinite complete set of conserved currents in involution. Quantum
mechanics is basically what is obtained from classical mechanics by
"quantization"; the fact that this process is not an isomorphism but a more
artistic kind of correspondence (the problem of ordering of operator products)
did not affect the inference of quantum integrability via quantization from
its classical counterpart. The best known illustration is the quantum analog
of the Kepler problem i.e. the hydrogen atom. In this case the conservation
laws which lead to integrability can be elegantly presented in terms a
spectrum-setting $O(4,2)$ group symmetry. Anomalies which could prevent
conservation laws to be inherited from their classical counterpart usually
need the presence of infinitely many degrees of freedom and hence occur
predominantly in quantum field theory (QFT).

There are many models of QFT which have remained outside the range of
Lagrangian quantization because no Lagrangian which fits them has been found;
in particular most of the so-called d=1+1 factorizing models, for which
explicit expressions for formfactors of quantum fields were constructed within
the bootstrap-formfactor program remained without a classical Lagrangian name.
Their given name refers to internal symmetries or to analogies with lattice
models. An example for such a situation is the scaling Z(N) Ising model in
\cite{BFK}, as the authors emphasize in the introduction of their paper.
Factorizing models constitute a nontrivial class of models with generally
noncanonical short distance behavior which owe their existence proof to
operator-algebraic methods. These methods are significantly different from the
quantum mechanical and measure-theoretical functional methods used by Glimm
and Jaffe \cite{G-J} in the 60s for establishing the existence of certain
canonical (superrenormalizable) models in the Lagrangian setting; in fact the
new method is based on \textit{modular localization} \cite{Lech2}.

Among the few integrable QFT which allow a Lagrangian presentation, the
Sine-Gordon model is the most prominent. The first indication about its
integrability came from the famous quasiclassical observations on the
Sine-Gordon particle spectrum by Dashen-Hasslacher and Neveu \cite{DHN}. But
even in cases like this, where the Lagrangian quantization setting even
provides a renormalized perturbation series, the latter still carries the
stain of divergence of all QFT perturbation series; Lagrangian quantization
"baptizes" a model with a name from classical field theory and permits a
perturbative expansion, but it does not lead to a proof of the existence of a
QFT behind this formalism, so that the problem of mathematically controlled
approximations cannot even be formulated.

Despite the observational success of the lowest terms in powers of the
coupling strength, one does not even know whether the power series is at least
an asymptotic approximand in the limit of vanishing coupling; the numerical
successes of renormalized QED and the subsequent observational achievements of
the standard model have no direct bearing on the mathematical consistency of
what is being approximated. Compared with other areas of theoretical physics
this is quite unique and shows, that in spite of almost 90 years which passed
since it was discovered, QFT still is not anywhere near its closure.

Recent insight into integrable d=1+1 factorizing model did not result from
refinements of the classical parallelism of the Lagrangian- or functional-
quantization setting. Rather it has been revealed through representation
theory, a method which was first introduced in 1939 by Wigner as a means to
obtain an intrinsic systematic classification of wave function spaces of
relativistic particles instead of having to cope with an ever-growing
confusing zoo of field equations, many leading to equivalent descriptions. In
the context of QFT, representation theoretical methods were first used in
connection with current algebras and chiral conformal QFTs. The discovery of
integrable QFTs with a particle interpretation started with
Dashen-Hasslacher-Neveu's \cite{DHN} quasiclassical observations about the
Sine-Gordon particle spectrum and took an interesting turn after it was
realized that it could be viewed as an exact realization of the "nuclear
democracy" particle spectrum which originated in the bootstrap setting of
scattering functions in d=1+1 \cite{STW}. The second important step was the
discovery of the representation theoretical bootstrap-formfactor setting
\cite{K-W}. This project reached its present perfection after it was realized that:

\begin{itemize}
\item Unitary and crossing symmetric elastic two-particle scattering functions
which obey the Yang-Baxter consistency relations can be classified
\cite{BKTW}\cite{K} and lead to combinatorial factorization formula for
n-particle elastic S-matrices. Zamolodchikov's formal algebraization added a
useful tool \cite{Zam} to the implementation of this (originally analytic)
classification project.

\item The bootstrap-formfactor program associates to each such scattering
function\footnote{In the presence of backward scattering and/or inner symmetry
indices the scattering function is a matrix function which fulfills the
Yang-Baxter equation \cite{Ba-Ka}.} explicitly computable formfactors of local
covariant fields from the local equivalence class (Borchers class) of
(composite) fields. The relation of the scattering function to an associated
QFT is unique (uniqueness of the solution of the inverse scattering problem
within the bootstrap-formfactor setting).

\item The creation and annihilation operators of the Zamolodchikov-Faddeev
(Z-F) algebra turned out to be the Fourier components of covariant
vacuum-polarization-free generators (PFGs) of interacting wedge-localized
algebras \cite{Sch1}\cite{Sch2}. They are special objects within the theory of
"modular localization" which permits to "emulate" wedge-localized products of
free fields \textit{inside the associated wedge-localized interacting algebra}.

\item The action of translations on a wedge-localized algebra together with
that of the modular reflection (in d=1+1 the TCP operation) generate a net of
right and left directed wedge algebras whose double cone intersections are
compactly localized algebras which act cyclic and separating on the vacuum
\cite{Lech1}\cite{Lech2}. In these papers the new constructive setting based
on modular localization obtained its first mathematical formulation. Combined
with the control of phase space degrees of freedom in the form of
\textit{modular nuclearity,} this approach led to the first existence proof of
QFTs with non-canonical short distance behavior within the algebraic setting
of QFT.

\item A sharp division between \textit{temperate} and \textit{non-temperate}
vacuum-polarization-free generators (PFGs) of wedge-localized algebras gives
rise to a dichotomy of integrable and non-integrable QFTs.
\end{itemize}

The integrable PFGs with their Z-F algebraic structure, apart from being
nonlocal (i.e. wedge- instead of point- local), relate particles and fields in
the standard way. Non-integrable PFGs on the other hand have translation
non-invariant domain properties which are radically different from those which
one meets in the standard formulation of fields or even with operators used in
nonlocal or noncommutative extensions \cite{BBS}. The much weaker relation of
particles with localized algebras and fields resulting from non-temperate PFGs
accounts for the difficulties one faces in studying non-integrable models of
QFT. It explains in particular why even 8 decades after its discovery, a
mathematically controlled construction of non-integrable models of QFT
remaines an open problem \cite{E-J}. Although our presentation in section 5
will not solve these problems, it does try to place them into sharper focus
which may be useful for future constructive attempts.

The DHR superselection theory \cite{Haag}, which constructs a full QFT from
its local observables algebras, leads to a different kind of "partial"
integrability which is of a more kinematical kind. In theories which are
non-integrable in the previous (dynamical) sense, the superselection strucure
of observable algebras can be described in terms of an exactly computable
combinatorial algebraic structure, even though the observable algebra itself
remains non-integrable. With some hindsight one can find out which integrable
algebraic superselection structure belongs to a specific non-integrable net of
local observable algebras. Since Lagrangian quantization leads directly to the
perturbation theory of the full field algebra (whose fixed-point algebra under
a compact inner symmetry group action defines the observable algebra), this
distinction between observables and their superselection structure is mainly
of conceptual, but hardly of practical interest.

The situation becomes more interesting for conformal QFT, in which case the
separation of observables from the superselection charge-carrying fields
arises in a natural way through the Huygens principle, according to which
observables commute for space- \textit{and time-like separations}. In the
context of QFT this limits the notion of observable algebras to algebras which
are generated by pointlike fields with integer scale dimension, whereas fields
which carry anomalous dimensions are considered as superselected charge
carriers similar to the DHR superselection of massive QFT. This attaches to
the anomalous dimensions a "Huygens superselection" aspect.

Indeed, the anomalous scale spectrum appears in the spectrum of the center of
the universal covering of the conformal group, and unlike the anomalous spin
spectrum in the Bargman-Wigner representation theory of the Lorentz group in
d=1+2, it is accompanied by the geometric covering of the (compactified)
Minkowski space. The latter consists of infinitely many "heavens" above and
"hells" below \cite{Lu-Ma} and hence the anomalous dimensions which control
the short distance behavior also show up in phase factors which arise from the
\textit{conformal rotation} in passing from one to Minkowski spacetime copy to
the next. \ \ 

In an appropriately extended terminology the anomalous dimensions together
with the braid-permutation group $\mathbf{BP}_{\infty}~$\cite{Fenn} structure
belong to the kinematical aspects of a conformal theory. In d=1+1 the
conformal group and the observable algebra split into right and left parts and
the $\mathbf{BP}_{\infty}~$group simplifies and becomes the braid group
$\mathbf{B}_{\infty},$ whereas the observable algebras (current,
energy-momentum) reduce to well-studied infinite Lie-algebras. The
combinatorial braid group algebra has interesting connections with Vaughn
Jones \cite{Jones} mathematical work on subfactors. In the last section this
situation will be presented in more details.

The natural setting for the modular theory is that of \textit{local quantum
physics} (LQP) \cite{Haag} which appeared in a rudimentary form already in
1957 in Haag's first attempt to formulate QFT in an intrinsic way \cite{Lille}
i.e. without leaning on the parallelism to the classical world of Lagrangian
fields. This formulation competed with the (at that time already existing)
framework by Wightman \cite{Wight} in which for the first time quantum fields
were identified with operator-valued Laurent Schwartz distributions. The
understanding of the inherently singular behavior of quantum fields as
compared to their classical counterparts was the key for "taming" the
ultraviolet divergence problem, which before almost caused the abandonment of
QFT. Haag's LQP approach attributes to these fields the role of (singular)
generators of a net of localized operator algebras; another helpful analogy is
that to coordinatizations in geometry.

Both settings were strongly influenced by Wigner's 1939 particle
classification in terms of group representation concepts for the inhomogeneous
Lorentz group (the Poincar\'{e} group $\mathcal{P)}$ as an alternative to the
quantization of classical field equations. Haag's basic idea was simple and
almost naive: measurements of local observables in a spacetime region
$\mathcal{O}$ which have a certain duration in time (the duration of the
activation of a particle counter) and a spatial extension within $\mathcal{O}$
should be members of an ensemble of operators forming an operator algebra
$\mathcal{A(O}).$ An experimentalist does not have to know the internal
structure of his particle- and radiation- counters, his only means to increase
precision is to improve their spacetime localizing sensitivity as well as
using several of them in coincidence and anti-coincidence arrangements. For
extracting scattering data from QFT it is not necessary to know the detailed
properties of an individual $\mathcal{O}$-localized observable, the
information that it belongs to a localized ensemble $\mathcal{A(O})$ and to
identify its superselected charge suffices \cite{Haag}.

The first test of this idea came from the derivation of the
Lehmann-Symanzik-Zimmermann (LSZ) scattering formula from the large-time
behavior of operators used in scattering theory in which only their
affiliation to a localized ensemble is used; differences between individual
operators in $\mathcal{A(O})$ only show up in adjustable numerical
normalization factors of their asymptotic limits (the \textit{insensitivity of
the S-matrix against local changes} \cite{Haag}). This viewpoint of
\textit{placing ensembles in form of localized operator algebras into the
center stage} has shown its soundness in numerous special situations; last not
least it accounts for the fact that there is no necessity for adding the
concept of probability to QFT since, in contrast to Heisenberg's QM to which
Born had to add the probability interpretation for individual events, the
thermal aspect together with its associated stochastic probability is
intrinsic in QFT. In fact it is a consequence of the mathematical description
of causal localization in the form of \textit{modular localization} which is
the foundational principle of LQP.

Born's spatial localization in terms of projectors from the spectral
decomposition of the selfadjoint position operator leads to a localization for
which the spatial restriction of the vacuum state (in the second quantized
description of Schr\"{o}dinger's QM) to the observables inside a localization
region does not acquire any additional property beyond those which it had as
an expectation value on the global observables. This is radically different in
LQP where the restriction of the vacuum to a spacetime localized algebra
$\mathcal{A(O})$ acquires a thermal manifestation which the \textit{global
vacuum} as an expectation value \textit{on all observables} did not have
\cite{E-J}. In mathematical terms, the restriction of the pure global vacuum
to localized variables is an impure thermal state to the causally closed
"world" associated with a local subalgebra of observables. This phenomenon is
best described in the setting of modular operator theory which among other
things attaches a \textit{modular Hamiltonian} $H_{mod}~$which is uniquely
determined by the pair ($\mathcal{A(O)},\Omega$)\ where $\Omega$ denotes the
global vacuum. The restricted vacuum is a thermal state which obeys the
KMS\footnote{The KMS relation is an anytic relation which is fulfilled by
tracial Gibbs states. It survives in the thermodynamic limit when the tracial
characterization is lost \cite{Haag}.} relation with respect to $H_{mod}.$
This property shows most clearly the conceptual rift between QM and QFT.

In the \textit{Einstein-Jordan conundrum} Einstein and Jordan \cite{E-J} came
close to this insight\textit{; }the observation that Jordan's subvolume
fluctuations in QFT are indistinguishable from Einstein's thermal black body
system as a consequence of "localization-thermality" which results from the
restriction of the pure vacuum state could have clinched the conundrum; but
the situation in 1925 was too much under the spell of the conceptual
revolution caused by the freshly discovered QM to permit the perception of
this subtle difference; so its full understanding had to await more than 8
decades. The phenomenon of vacuum polarization, which was discovered some
years later by Heisenberg, has a strong relation to the E-J
conundrum\footnote{In fact in a private correspondence Heisenberg challenged
Jordan to account for a logarithmic divergence resulting from the vacuum
polarization cloud at the endpoints of his localization interval \cite{Du-Ja}
\par
.
\par
. $~$} since the formation of vacuum polarization clouds at the boundary of a
localization region (leading to area proportionality of \textit{localization
entropy}) and the thermal property of the reduced vacuum state are two sides
of the same coin.

There can be no doubt that Einstein, who had a livelong philosophical problem
with Born's assignment of probabilities to individual events in QM, would have
embraced the the ensemble probability from thermalization of a subvolume
restricted vacuum state in QFT. After all it was the subvolume fluctuations in
the statistical mechanics at finite temperature which led him to the
corpuscular aspect of light \cite{E-J}. But time was not yet ripe to
understand that subvolume restriction in QFT in constrast to QM lead to a loss
of purity of the restricted vacuum state which results in thermal behavior
without a "heat bath" being present

The existence of a thermal manifestation of causal localization is an
unavoidable consequence of Haag's quantum adaptation of the Faraday-Maxwell
"Nahewirkungsprinzip" \cite{Haag} (together with Einstein's refinement of
causality in Minkowski spacetime). In fact causal locality and its more recent
mathematical formulation as \textit{modular localization} became the
cornerstone of Haag's local quantum physics (LQP) and the thermal
manifestation of the subvolume restricted vacuum is a consequence.

The perception of the presence of this intrinsic ensemble probability could
have vindicated Einstein's philosophical resistance against Born's
probability, so that many physicists nowadays would not think of Einstein's
reluctance as being the stubborn resistence of an old man against the tides of
the time. Einstein may have even accepted Born's assignment of probability to
individual events if it were possible to show that within a better conceptual
understanding of the limiting relation of QM with the more fundamental local
QFT, the probabilistic manifestation of a localized ensemble passes to Born's
quantum mechanical probability assigned to individual\footnote{QM lacks the
ensemble aspect which results from modular localization; The Born-localization
resulting from the projectors of the spectral decomposition of the has does
not have an intrinsic ensemble-probabiliy interpretation; Born had to
postulate it..} events even though causal localization and vacuum polarization
disappear in such a limit.

As most ideas which do not result from the extension of an already existing
formalism but rather emerged from philosophical contemplations about physical
principles, their content is usually more subtle than that of the supporting
intuitive arguments. Already for the localization inside a non-compact
spacetime region as big as a Rindler wedge only a uniform acceleration in the
wedge direction can keep a particle counter (observer) inside a non-compact
wedge region (the Unruh Gedankenexperiment \cite{Unruh}); and when it comes to
the localization inside a compact causally closed region (e.g. the double cone
completion of a spatial ball) it becomes impossible in models with a mass-gap
to visualize the realization of modular localization in spacetime
geometric/physical terms since the region preserving modular group becomes an
abstract automorphism of the localized operator algebra.

The thermal manifestation of modular localization is an unavoidable
consequence of Haag's adaptation of classical causality (Maxwell equations) to
the requirements of quantum theory, and hence it is fully present in QFT. The
impurity of the state resulting from the restriction of the pure global vacuum
state to the ensemble of observables contained in localized (without loss of
generality causally closed) algebra and its description in terms of a KMS
state associated with a "modular Hamiltonian" is a mathematical fact,
notwithstanding the difficulties to find direct observational
verifications\footnote{There is an ongoing discussion whether thermal
radiation effects in Unruh situations can be seen in subnuclear laboratory
experiments.} which underlines the subtlety of causal localization in QT.

It is not important that physical principles can be directly verified in terms
of existing hardware or whether their intuitive support under closer
conceptual scrutiny appears somewhat metaphoric; important is primarily the
mathematical precision of the formulation and the wealth of the theoretical
and observational consequences. For the Unruh Gedankenexperiment, the
\textit{dependence of the state of the observer} attributes a certain aura of
\textit{fleetingness}, but this is to some extend overcome if the horizon is
an \textit{intrinsic property of the spacetime metric in form of an event
horizon}. The curvature contained in the spacetime metric does not create the
thermal aspects of Hawking radiation, it rather replaces the
observer-dependent fleeting aspect of the causal horizon by a more robust
observer-independent \textit{event horizon} and in this way favors the
macroscopic detection of localization-thermality in astrophysical observations.

In contrast to the volume-proportional heat bath entropy, the
\textit{localization entropy~of quantum matter }is proportional to the
dimensionless area $A/\varepsilon^{2}\rightarrow\infty,~\varepsilon
\rightarrow0$ where $\varepsilon$\textit{~}is the surface "roughness" i.e. the
thickness of the layer which is conceded to the attenuation of the vacuum
polarization cloud. In the Bekenstein conjecture the finite black hole entropy
(formally obtained by replacing the parameter $\varepsilon~$by the Planck
distance) refers to the hypothetical degrees of freedom of a future quantum
theory of gravity. We believe that this is related to 't Hooft's brickwall
picture \cite{brick}. Such a surface divergence in the limit of sharp
localization appeared for the first time in Heisenberg's treatment of vacuum
polarization caused by localization of conserved (dimensionless) charges
\cite{E-J}.

It is not surprising that the subtlety of the principle of causal localization
has led to misunderstandings in particular in string theory (last section).
Although the correction of deep errors stemming from misunderstandings is a
powerful method to highlight the importance of a principle, we will not follow
such a path here and instead refer to other publications \cite{cau}%
\cite{foun}\cite{response}.

There remains the question why these important localization properties of QFT
have not been seen already long ago. The answer is that they did not play an
important role in the Tomonaga-Schwinger-Feynman-Dyson discovery of
renormalized perturbation theory; the essential step from the old perturbation
theory (in textbooks by Heitler and Wentzel), which had insurmountable
problems with vacuum-polarization, to the modern formulation was the
\textit{implementation of a covariant formulation}. Covariance under the
Poincar\'{e} automorphisms of Minkowski spacetime is closely related to causal
localization, but it generally leads to stronger results.

The sections 2 and 4 prepare the ground for "modular localization" which is
used in section 5 for a characterization of integrability/non-integrability in
terms of properties of generators of wedge-localized operator algebras. Since
modular localization is a quite subtle recent concept, and not every reader is
prepared to invest in conceptual/mathematical ideas without some indication
about their physical relevance for important open problems of particle
physics, it may be helpful to indicate what one hopes to achieve from a
reformulation of the BRST gauge theory in terms of string-localized
vectorpotentials (section 3); this new view may even lead to revisions of the
Higgs issue.

\section{The modular localization approach of QFT}

There are two routes to modular localization, a mathematical access and a more
physical-conceptual path. The mathematical route starts from the
Tomita-Takesaki modular theory of operator algebras and makes contact with QFT
by applying this to Haag's LQP algebraic formulation of QFT in terms of
spacetime-indexed nets of operator subalgebras \cite{Haag}. An important step
was the recognition by Bisognano and Wichmann \cite{Bi-Wi}\cite{Mund} that the
abstract Tomita-Takesaki modular group $\Delta_{\mathcal{A}}^{it}$ and the
modular reflection $J$ acquire a direct geometric-physical meaning in terms of
particle physics concepts in case of a wedge-localized operator
subalgebra\footnote{A general wedge results from $W_{0}=\left\{  \left\vert
x_{0}\right\vert <x_{1}\right\}  ~$by applying Poincar\'{e} transformations.}
$\mathcal{A(}W),$ whereas the modular objects for compact localized algebras
$\mathcal{A(O})$ can be in principle determined by representing the region of
interests (rather its causal completion) and the associated algebras using
intersections of wedges and wedge-localized algebras \cite{Summers}.

Modular theory began in the middle of the 60s as a joint venture between
mathematics and physics when, at a conference in Baton Rouge \cite{Borch},
mathematician interested in operator algebras (Kadison, Tomita, Takesaki) met
physicists (Haag, Hugenholz, Winnink) who just had finished work on an
intrinsic formulation of statistical \textit{quantum mechanics of open
systems} which avoids the non-covariant box-quantized Gibbs states by starting
directly in the thermodynamic infinite volume limit \cite{Haag}.

In this work they used an older analytic trick which Kubo, Martin and
Schwinger to avoid computing Gibbs traces. In the open system setting this
acquired a fundamental conceptual significance; in this way the KMS property
became part of the joint mathematics/physics heritage, combining modular
theory of operator algebras (where it led to Connes famous classification work
about von Neumann algebras) with statistical mechanics of open systems.
Whereas the box-quantized thermal Gibbs states always stay within the setting
of the standard (type I$_{\infty})~$quantum mechanical algebras $B(H)$ of all
bounded operator on a Hilbert space, the thermodynamic limit\footnote{The
tensor factorization of type I$_{\infty}~$"thermofield theory" breaks down and
the algebra changes its type.} changes the algebraic type into what afterwards
in Connes classification was called the unique hyperfinite type III$_{1}$ von
Neumann factor algebra.

At the beginning of the 60s Araki \cite{Haag} had already shown that such
algebras, which have quite different properties from those met in QM, occur in
the form of local operator algebras in QFT. Together with the statistical work
on open systems this was suggestive of an conceptual connection of thermal
behavior and localization but at that time this remained unnoticed. It came as
quite a surprise when a decade afterwards Bisognano and Wichmann \cite{Bi-Wi}
discovered that the \textit{monad} \footnote{The short name which we will use
for the (up to isomorphism unique) befor-mentioned operator algebra. Besides
the standard algebra $B(H)$ of all bounded operators this the only type of
operator algebra which one encounters in continuous quantum systems
\cite{Jakob}.}, realized as a wedge-localized subalgebra $\mathcal{A}(W),$ has
modular data which have well-known physical/geometrical significance,
including the KMS property. As a result it became clear that localization,
thermalization, and the generation of vacuum polarization clouds are
inexorably intertwined.

Using the representation theoretical access to modular localization, we begin
our presentation with extracting modular objects from Wigner's classification
of irreducible positive energy representations of the Poincar\'{e} group. This
concept was not available to Wigner who realized that the Born localization
based on the position operator to the relativistic inner product (the
\textit{Newton-Wigner localization}) was not the right concept which could his
particle representation theory with QFT. This may explain why Wigner, after
his important contribution to QFT immediately after its discovery, maintained
a lifelong critical distance with respect to its later developments. Whereas
the Born localization is extrinsic\footnote{Born localization entered QM
through his famous probabilistic interpretation of (the Born approximation of)
the scattering amplitude i.e. the cross section. This was afterwards extended
to the position operator and its associated wave functions. .} to QM, the
modular localization and its thermal-probabilistic aspect is intrinsic, i.e.
it only uses concepts from the representation theory of the Poincar\'{e}
group. For matters of notational simplicity we restrict our presentation to
the case of a scalar massive particle.

It has been realized, first in a special context in \cite{Sch1} and then in a
general mathematical rigorous setting which covers all positive energy
representations in \cite{BGL} (see also \cite{Fa-Sc}\cite{MSY}), that there
exists a natural localization structure on the Wigner representation space for
any positive energy representation of the proper Poincar\'{e} group. The
starting point is an irreducible $(m>0,s=0)$ one-particle representation of
the Poincar\'{e} group on a Hilbert space $H$\footnote{Since positive energy
representations are completely reducible this works for all such
representations, not only irreducible ones.} of wave functions with the inner
product
\begin{equation}
\left(  \varphi_{1},\varphi_{2}\right)  =\int\bar{\varphi}_{1}(p)\varphi
_{2}(p)\frac{d^{3}p}{2p_{0}}%
\end{equation}
For other (higher spin, massless) representations the relation between the
momentum space wave function on the mass shell (or light cone) and the
covariant wave functions is more involved as a consequence of the presence of
intertwiners $u(p,s)$ which connect the unitary with the covariant
representations. Selecting a wedge region $W_{0}=\{x\in\mathbb{R}^{d}%
,x^{d-1}>\left\vert x^{0}\right\vert \}$ one notices that the unitary
wedge-preserving boost $U(\Lambda_{W}(\chi=-2\pi t))=\Delta^{it}$ commutes
with the antiunitary reflection $J_{W}$ on the edge of the wedge (i.e. along
the coordinates $x^{d-1}-x^{0}$). The distinguished role of the wedge region
is that it produces a commuting pair of boost and antiunitary reflection. This
has the unusual (and perhaps even unexpected) consequence that the closed,
antiunitary operator (the Tomita S-operator),%
\begin{align}
S_{W}  &  :=J_{W}\Delta^{\frac{1}{2}},~~S_{W}^{2}\subset1\\
&  since~~J\Delta^{\frac{1}{2}}J=\Delta^{-\frac{1}{2}}\nonumber
\end{align}
which is intrinsically defined in terms of Wigner representation data, is
\textit{involutive on its dense domain} and has a unique closure (unchanged
notation) with $ranS=domS.$

The involutivity means that the s-operator has $\pm1$ eigenspaces; since it is
antilinear, the +space multiplied with $i$ changes the sign and becomes the -
space; hence it suffices to introduce a notation for just one eigenspace%
\begin{align}
K(W)  &  =\{domain~of~\Delta_{W}^{\frac{1}{2}},~S_{W}\psi=\psi
\},~K(W)~closed\label{K}\\
&  J_{W}K(W)=K(W^{\prime})=K(W)^{\prime},\text{ }duality\nonumber\\
&  \overline{K(W)+iK(W)}=H,\text{ }K(W)\cap iK(W)=0\nonumber
\end{align}

It is important to be aware that, unlike QM, we are dealing here with real
(closed) subspaces $K$ of the complex one-particle Wigner representation space
$H$. An alternative is to directly work with complex dense subspaces
$K(W)+iK(W)$ (third line). Introducing the \textit{graph norm} in terms of the
positive operator$~\Delta,$ the dense complex subspace becomes a Hilbert space
in its own right. The second and third line require some more explanation: the
upper dash on regions denotes the causal disjoint (the opposite wedge),
whereas the dash on real subspaces stands for the symplectic complement with
respect to the symplectic form $Im(\cdot,\cdot)$ on $H.$

The two properties in the third line are the defining relations of what is
called the \textit{standardness property} of a real
subspace\footnote{According to the Reeh-Schlieder \cite{Haag} theorem a local
algebra $\mathcal{A(O})$ in QFT is in standard position with respect to the
vacuum i.e. it acts on the vacuum in a cyclic and separating manner. The
spatial standardness, which follows directly from Wigner representation
theory, is just the one-particle projection of the Reeh-Schlieder property.};
any standard $K$ space permits to define an abstract s-operator%
\begin{align}
S(\psi+i\varphi)  &  =\psi-i\varphi\label{inv}\\
S  &  =J\Delta^{\frac{1}{2}}\nonumber
\end{align}
whose polar decomposition (written in the second line) yields two modular
objects, a unitary modular group $\Delta^{it}$ and an antiunitary reflection
$J,$ which generally have however no geometric significance. The domain of the
Tomita $S$-operator is the same as the domain of $\Delta^{\frac{1}{2}},$
namely the real sum of the $K$ space and its imaginary multiple. In our case
this domain is determined solely in terms of Wigner group representation theory.

Coming from QFT, the complex $domS_{W}$ can be understood as the complex dense
space which is provided by projecting the dense Reeh-Schlieder domain
(obtained by applying fields smeared with W-supported test functions to the
vacuum \cite{Haag}) to the one-particle space, and the closed $K(W)$ space
results from projecting only Hermitian operators. The modular localization
approach provides a constructive access to QFT whithout quantization by
avoiding any parallelism with the less fundamental classical field
theory\footnote{In Jordan's terminology "without classical crutches"
\cite{Jor}.}. As will be demonstrated in section 4, this is quite easy in the
absence of interactions, but leads to new concepts which pose novel problems
in the presence of interactions (section 5). The $J~$transformed $K$-spaces
(\ref{K}) permit a more direct description in terms of symplectic complements
\begin{equation}
K^{\prime}:=JK=\{\chi|~Im(\chi,\varphi)=0,~all\text{ }\varphi\in K\}
\end{equation}

It is easy to obtain a net of K-spaces by $U(a,\Lambda)$-transforming the
K-space of a particular $W_{0}.$ A bit more tricky is the construction of
sharper localized subspaces via intersections
\begin{equation}
K(\mathcal{O})=%
{\displaystyle\bigcap\limits_{W\supset\mathcal{O}}}
K(W)
\end{equation}
where $\mathcal{O}$ denotes a causally complete smaller region (e.g.
non-compact spacelike cone, compact double cone). Intersection may not be
standard, in fact they could even be zero, in which case the theory allows
localization in $W$ (it always does), but not in $\mathcal{O}.$ One can show
that the intersection for non-compact \textit{spacelike cones} $\mathcal{O=C}$
is always standard for all positive energy representations \cite{BGL}.

Standardness for compact double cone regions $\mathcal{O=D}$ leads to
pointlike localized generating wave functions (wave-function-valued Schwartz
distributions). This applies to$\ (m>0,s)$ and to massless finite helicity
representations, whereas the Wigner massless infinite spin family with
$\mathcal{K(D)}=0,\ K(\mathcal{C})~$standard, requires semiinfinite spacelike
string generating wave functions. In the functorial relation between Wigner
wave functions and quantum fields this leads to pointlike/stringlike localized
generating free fields. The positive energy Wigner representations fall into 3
families: positive mass, zero mass finite helicity and zero mass infinite
helicity. Only the third class, for which the two-dimensional Euclidean little
group is faithfully represented, requires stringlike generating wave functions.

Since all states in QFT carry a unitary representation of the Poincar\'{e}
group which permits a (discrete or continuous) decomposition into irreducible
components, this closes the issue of modular state-localization. Modular
operator localization of free fields follows state-localization (section 4),
however modular localization of interacting operator algebras requires a
subtle refinement which accounts for new conceptual problems posed by
interactions (section 5).

For the explicit construction of the pointline free fields of arbitrary mass
and finite spin it is somewhat easier to follow Weinberg \cite{Weinbook} and
compute covariant \textit{intertwiners} which map momentum space creation and
annihilation operators into covariant fields. This is possible because in the
absence of interactions modular localization and covariance are equivalent
requirements and the differences between Wigner's "little group" and its
representation accounts fully for the differences in localization between the
three representation families.

Leaving out the string-localized infinite spin family, the result of the
"covariantization" associates one unitary (m,s) Wigner representation with an
infinite family of generating covariant spinorial wave functions
$\Psi^{(A,\dot{B})}$ whose spinorial undotted/dotted spinorial indices are
related to the physical spin s through the following inequalities\footnote{For
convenience of notation our spinorial indices are half of the standard ones.}%

\begin{align}
\left\vert A-\dot{B}\right\vert  &  \leq s\leq\left\vert A+\dot{B}\right\vert
,\text{ }m>0\label{line1}\\
s  &  =\left\vert A-\dot{B}\right\vert ,~m=0 \label{line2}%
\end{align}
One notices that in the zero mass case the vector representation
($A=1/2,B=1/2$) for s=1 and the ($A=1,B=1$)~for s=2 are missing i.e. precisely
those fields which correspond to the classic electromagnetic vectorpotential
and for s=2 to the metric tensor. These gaps in the massless case have
important physical consequences.

Wigner's representation theory for positive energy representations of the
Poincar\'{e} group combined with the calculation of intertwiners via
covariantization represents a completely intrinsic quantum path to free
fields. The passing from generating covariant wave function to covariant
quantum fields only requires to reinterpret the momentum space wave functions
$a^{\ast}(p)$ and their antiparticle counterpart $b(p)$ as canonical
creation/annihilation operators%
\begin{equation}
\Psi^{(A,\dot{B})}(x)=\frac{1}{\left(  2\pi\right)  ^{3/2}}\int\frac{d^{3}%
p}{2p_{0}}\{e^{ipx}u^{(A,\dot{B})}(p)\cdot a^{\ast}(p)+e^{-ipx}v^{(A,\dot{B}%
)}(p)\cdot b(p)\} \label{cov}%
\end{equation}
Here the dot stands for the summation over physical spin components and the
dependence on the spinorial component of the ($A,\dot{B})$ representations
(ranging over $2A+1$ respectively $2\dot{B}+1$ values) on the $\Psi$ and the
$u$, $v$ intertwiners have been omitted. The covariantization leading to the
intertwiners uses only group theory, it can be found in the first volume of
Weinberg's well-known book \cite{Weinbook}.

There is a subtle consequence of modular localization which one encounters in
the second ($m=0,s\geq1$) representation class of massless finite helicity
representations (the photon-graviton family). Whereas in the massive case the
relation of the physical spin $s$ with the formal spin in the spinorial fields
follows the angular momentum composition rules which leads to the spinorial
restrictions (\ref{line1}) \cite{Weinbook}, the zero mass finite helicity
family in the second line has a significantly reduced number of spinorial
descriptions. Different from classical Maxwell theory, where pointlike
vectorpotentials are perfectly acceptable (constrained) classical fields,
their quantum counterparts do not appear in the covariantized Wigner's list
(\ref{line1}).

The explanation of this dilemma, which also leads to its cure, is that the
loss of pointlike quantum potentials is the result of a \textit{clash between
the Hilbert space structure (positivity) and pointlike localization}. The
missing spinorial fields in (\ref{line2}), as compared to (\ref{line1}),
reappear after relaxing the localization from pointlike to stringlike
\cite{MSY}. Both kind of fields are singular limits of operators localized in
causally closed regions; pointlike fields in case of double cone localized and
semi-infinite stringlike fields in case of spacelike cone localized operators.
Once one allows stringlike covariant fields i.e. $\Psi^{(A,\dot{B})}(x,e)$
localized on spacelike half-lines $x+\mathbb{R}_{+}e,$ the full range of
spinorial realizations (\ref{line2}) is available. These generating free
fields are covariant and "string-local"%

\begin{align}
U(\Lambda)\Psi^{(A,\dot{B})}(x,e)U^{\ast}(\Lambda)  &  =D^{(A,\dot{B}%
)}(\Lambda^{-1})\Psi^{(A,\dot{B})}(\Lambda x,\Lambda e)\label{string}\\
\left[  \Psi^{(A,\dot{B})}(x,e),\Psi^{(A^{\prime},\dot{B}^{\prime})}%
(x^{\prime},e^{\prime}\right]  _{\pm}  &  =0,~x+\mathbb{R}_{+}e><x^{\prime
}+\mathbb{R}_{+}e^{\prime}\nonumber
\end{align}
although the Wigner representation itself (and its functorially related local
operator algebra, see next section) remains pointlike generated, since the
field strengths suffice for its generation. But for other purposes the
potentials are indispensable. (see below). The use of covariant string fields
also facilitates the construction of finite "gauge bridges" (using the
terminology of gauge theory) between two matter fields with opposite charges.

The third Wigner family, the \textit{infinite spin representations, }resisted
all attempts to understand their localization properties for a very long time.
Its local generators have no analog in classical Lagrangian field theory and
also did not appear in Weinberg's intertwiner formalism \cite{Weinbook}. Only
after the concept of modular localization was applied to Wigner's
representation theoretical construction \cite{BGL}\cite{MSY}, it became clear
that this is a case of a string-like generated Wigner representation. The
differences in localization properties can be traced back to the different
representations of the Wigner "little groups". For m=0 representations the
little group is the two-dimensional Euclidean group, but only for the
\textit{faithful} $E(2)$ representation the non-compactness of the group makes
itself felt in the necessity to introduce stringlike generators. The
interwiners do not have spinorial indices, instead they depend on a string
direction $e$ and the dot in (\ref{cov}) stands for an infinite sum reflecting
the infinite dimensional nature of the $E(2)$ representation space.

The conventional description of vectorpotentials, which one obtains from
quantization of their classical counterpart, maintains the pointlike
formalisms at the expense of Hilbert space structure. The application of the
BRST formalism gives the physically correct results only for gauge invariant
quantities, which are automatically pointlike localized \cite{nonlocal}. But
the physical electric charge-carrying operators are known to allow no
localization which is sharper then an arbitrary thin spacelike cone with
stringlike generating fields which remain outside the BRST formalism. The use
of the string-localized vectorpotentials exposes the origin of their
string-like localization by coupling the quantum matter to string-like
vectorpotentials. The result is that their stringlike localization is exported
to the matter field, which in zero order perturbation was point-like
localized. Whereas the vectorpotential continues to lead to pointlike field
strengths, there is no linear operation which undoes the string localization
of the charge-carrying field.

Even in the case of free fields the use of stringlike vectorpotentials
protects against incorrect application of the gauge formalism. A well-known
illustrations is the Aharonov-Bohm effect in QFT\footnote{The standard A-B
effect refers to QM in an external magnetic field.}. It is not necessary to
use vectorpotentials, but if one decides to use them it is important to work
with the string-localized potential since the pointlike indefinite metric
potential (the Feynman gauge) gives a wrong answer \cite{charge}%
\cite{nonlocal}.

Stringlike massless higher spin "potentials" have a better short distance
behavior than their associated field strength (whose short distance dimension
increases with $s)$; in fact in all cases one finds potentials with $d_{sd}=1$
which is the smallest dimension allowed by Hilbert space positivity and also
the largest for satisfying the power counting renormalization requirement in
up to quadrilinear polynomial couplings. As a result there are candidates for
renormalizable interactions for any spin. Of course renormalization theory is
more than power-counting; one also has to show that an extension of the
Epstein-Glaser iterative implementation of causal commutativity can be
implemented for stringlike fields.

Stringlike localized covariant fields can also be constructed in massive
higher spin theories. In this case the pointlike potentials exist, but their
dimensions increases with $s$.~The string-like description and the BRST
formalism both use massive potentials with $d_{sd}=1.~$Even the relation to
the pointlike physical Proca potential is formally similar \cite{Rio}%
\begin{align}
&  A_{\mu}(x,e)=A_{\mu}^{P}(x)+\partial_{\mu}\phi(x,e),~A_{\mu}^{BRST}%
(x)=A_{\mu}^{P}(x)+\partial_{\mu}\phi^{S}(x)\label{proca}\\
&  \left\langle A_{\mu}(x,e)A_{v}(x^{\prime},e^{\prime})\right\rangle
=\frac{1}{\left(  2\pi\right)  ^{3}}\int\frac{d^{3}p}{2p_{0}}%
e^{-ip(x-x^{\prime})}\{-g_{\mu\nu}-\nonumber\\
&  -\frac{p_{\nu}p_{\nu}(e,e^{\prime})}{(pe)_{-i\varepsilon}(pe^{\prime
})_{-i\varepsilon}}+\frac{p_{\mu}e_{\nu}}{(pe)_{-i\varepsilon}}+\frac{p_{\mu
}e_{\nu}^{\prime}}{(pe^{\prime})_{-i\varepsilon}}\}\nonumber
\end{align}
where the $\varepsilon$-prescription refers to the way in which the real
boundary in $e$ has to be approached. The difference in (\ref{proca}) is that
the scalar St\"{u}ckelberg field $\phi^{S}(x)$ has the opposite metric (it is
related to $A_{\mu}^{BRST}(x)$ via the s-operation of the BRST formalism
$sA_{\mu}^{BRST}=\partial_{\mu}u,~s\phi^{S}=u$). The use of the directional
derivative $\partial_{e}=\sum e^{\alpha}\partial_{e^{\alpha}}$ leads to an
even stronger formal connection%
\begin{align*}
\partial_{e}A_{\mu}(x,e)  &  =\partial_{\mu}v(x),~\partial_{e}\phi(x,e)=v(x)\\
sA_{\mu}^{BRST}(x)  &  =\partial_{\mu}u(x),~s\phi^{S}(x)=u(x)
\end{align*}
where $v,~$the counterpart of $u,$ is pointlike. This stringlike description
is reminiscent of Mandelstam's attempt \cite{Mand} to avoid indefinite metric
by expressing the dynamics in terms of field strength only. Indeed a
stringlike potential is uniquely determined by the field strength and a
spacelike direction $e;$ but the introduction of a separate operator $A_{\mu
}(x,e)$ which fluctuates in both $x$ and $e$ is preferable, because the
improvement of the short distance $x$-fluctuation which is crucial for
renormalization, and its prize in terms of the appearance of infrared
fluctuations, is placed into evidence.

The more surprising consequences of modular localization are certainly those
which appear in the title of this paper; they will be presented in section 5.

\section{Some expected consequences}

The stringlike reformulation of gauge theories leads to a setting which has
some formal similarities with the BRST formalism\footnote{Concerning the
application of the BRST formalism to massive vectormeson we follow Scharf's
book \cite{Scharf}.} without suffering from its limitations. Its use for
massive vectormesons relates their string-localized description, which has the
mild short distance behavior needed for renormalizability, with the standard
physical description in terms of pointlike Proca field, similar to the
relation between the pointlike indefinite metric BRST vectorpotential with the
Proca field (\ref{proca}). There are reasons to believe that such a relation
has its multiplicative counterpart in a relation between a stringlike
interacting matter field (unavoidable as the result of its coupling with the
string-localized potential) and a multiplicatively corresponding pointlike
matter field which together with the pointlike vectormeson leads to a better
description (\textit{after} the issue of renormalization has been settled).
Both set of fields belong to the same theory, they are relatively local
generating fields acting in the same Hilbert space. The pointlike fields
permit to present the result in the standard way, but in contradistinction to
interactions between fields with $s<1,$ they are \textit{not} suitable for
staying within the power-counting limit required by renormalization. In
contrast to the BRST description one does not need Krein spaces; the role of
the short distance-improving indefinite metric $A_{\mu}^{BRST~}$is now taken
over by the stringlike vectorpotential which, apart from being
string-localized, is expected to belong to the same QFT as the Proca field.

The conceptual advantage of the stringlike formulation emerges in a much
stronger form in the zero mass limit. The Proca potential cannot have a zero
mass limit since a physical (Hilbert space) pointlike zero mass
vectorpotential does not exist. In fact there are two different BRST settings
\cite{Scharf}; one for massive vectormesons, in which case the BRST formalism
is only used on the level of the free asymptotic fields, and the other for the
massless gauge theories, where the BRST s-operation has a perturbative
dependence of the coupling strength. In the stringlike formulation everything
is under one conceptual roof, the pointlike potentials are only lost in the
zero mass limit. Computation with stringlike fields should be done for massive
vectormesons; one hopes that the connection of string-localization and
infrared properties is more amenable than in the massless limit.

For the matter fields, the main change between the stringlike matter fields
and its would-be multiplicative related pointlike counterpart is expected to
occur in the long-distance regime, whereas for short distances there should be
essentially no change. For nonabelian zero mass models one naturally expects
that the linear field strength remains stringlike and only appropriate
composites maintain their pointlike localization i.e. the logic of classical
gauge invariance is replaced by the identification of a pointlike generated
observable substrate embedded in a string-generated QFT.

Approaching QED from the massive string-localized side has the advantage that
physical charged fields (infraparticle fields), which have been known for a
long time to acquire a non-compact localization \cite{Haag}, are now part of
the perturbative formalism. Whereas in the existing formulation they only
appear as a computational device in a prescription for computing on-shell
photon inclusive scattering cross sections, their expected new role is to
represent electric charge-carrying physical "infraparticle" fields with admit
(off-shell) correlation function. The largest gain of insight from
string-localization is expected to come from approaching Yang-Mills theories
via massive self-interacting vectormesons. Whereas the physical origin of
infrared singularities of the unphysical pointlike matter fields in the BRST
setting remains unexplained (or is blamed on a not yet understood
nonperturbative regime), the stringlike formulation may reveal a different story.

The string-localization plays a pivotal role for the understanding of the
off-shell infrared divergencies of self-interacting massive vectormesons in
the zero mass limit. In the nonabelian case only color-neutral composites of
the field strengths are expected to maintain their pointlike localization in
the massless limit. It is perfectly conceivable that gluon strings exist in
the Hilbert space but that the expectation of the energy-momentum operator in
a state created from the vacuum by a gluon operator leads to a diverging
global energy.

This could have interesting consequences for the asymptotic freedom issue and
for the confinement idea. A beta-function computed in the dimensional
regularization formalism without the Callan-Symanzik equation from where it
originates remains incomplete. One needs to represent the infrared-divergent
massless Yang-Mills theory as the massless limit of a perturbatively
accessible massive theory\footnote{The prototype illustration is the
derivation of the C-S equation of the massive Thirring model. In this case the
vanishing of beta to all orders insures the existence of the massless limit as
a conformal QFT \cite{Lo-Go}.} in order to derive a C-S equation. The existing
derivation of asymptotic freedom is tantamount to a consistency argument: the
inverse sign of the beta function is consistent with the expected perturbative
short distance behavior of a nonexisting renormalized perturbation theory.

The reformulation Y-M interaction in terms of stringlike vectorpotentials
permits to question this scenario; massive stringlike selfinteracting
vectormesons may have a massless limit, a situation which cannot be achieved
in the pointlike BRST formalism. Assuming that Callan-Symanzik equations also
can be established in the stringlike setting, such a step would combine
massive $s=1$ fields and their massless limit under one roof, just as in case
of interactions with $s<1~$for which this limit for pointlike fields. Such a
scenario, if it can be made to work, could change the whole perspective on QFT
and its use in particle physics. String-localized potentials with their
string-improved short distance dimension $d_{sd}=1$ are potential candidates
for renormalizable interactions between \textit{all} $(m,s\geq1)$ quantum fields.

The second area which could suffer significant modifications is the issue
around the Higgs boson. From a "Schwingerian" point of view what remains of
the complex matter field in the screened "phase" of scalar QED (which is a
three-parametric renormalizable QFT in terms of $m_{sc},g_{sc},e)~$is a real
scalar matter field, after its remaining real partner participated in the
conversion of the photon into a massive vectormeson. The screened phase of QED
is one in which the \textit{Maxwell charge} (related to the identically
conserved current in $\partial^{\mu}F_{\mu\nu}=j_{\nu})$ is "screened" i.e.
the integral over $j_{0}$ vanishes. Schwinger's screening
idea\footnote{Schwinger's attempt to find a perturbative realization in spinor
QED failed and he instead used two-dimensional QED which became known as the
"Schwinger model". He was apparently not aware that by replacing spinors by
complex scalar fields he could have had a perturbative realization of his idea
\cite{Swieca}.} is a QFT extension of Debye screening in QM, except that in
QFT this is more radical (change of particle spectrum). This idea was later
backed up by a theorem due to Swieca \cite{Sw}\cite{Swieca} which states that
this situation permits precisely two alternative realizations: either the
charge is $\neq0,~$in which case the mass-shell is converted into the milder
cut singularity of infraparticle, or there is a mass gap which leads to a zero
charge (the screening phase).

Its discoverers Higgs \cite{Higgs} and \cite{Eng}\cite{Gur} presented the
result as a two step process in which a spontaneous symmetry breaking is
followed by a transfer of the massless Goldstone boson to the vectorpotential
which, as a result of gaining an additional degree of freedom, becomes a
massive vectormeson. This idea was helpful from a computational viewpoint
(since the perturbative Goldstone spontaneous symmetry breaking mechanism was
well-known at the time of Higgs). The result is the same as that computed from
Schwinger-Higgs screening except that there is no intermediate spontaneous
symmetry breaking. Unfortunately the terminology "broken gauge invariance" led
to a rather widespread incorrect understanding of the Higgs mechanism. There
is \textit{no symmetry breaking}, unless one wants to consider the appearance
of odd terms in the self-interaction of the remaining screened (real) scalar
field as a spontaneous $\mathbb{Z}_{2}$ symmetry breaking.

The renormalization theory of massive vectormesons appeared first in the form
of massive QED \cite{L-S}. Some loose ends of that treatment concerning the
passing from the indefinite metric description to the unitary gauge were
overcome in the BRST formalism. In recent work \cite{Due}\cite{Ga} which
extends Scharf's \cite{Scharf} pathbreaking work\footnote{Besides his
formulation of operator gauge invariance Scharf uses only the intrinsic logic
of the BRST formalism.} on the BRST setting of massive vectormesons, the
absence of any symmetry breaking is emphasized \textit{against the mainstream
view}. To the extend that the mainstream pays attention to this work, also the
interest in Schwinger-Higgs screening and the screening theorem by Swieca
(which had a foundational impact on the development of LQP\footnote{It led
Buchholz and Fredenhagen to discover the connection of spectral gaps with the
stringlike generating property of superselected charge-carrying fields (as the
only possibility which replaces the pointlike localization in case of
non-existence of pointlike generators) \cite{Haag}.} but never made it into
the mainstream) may receive the attention it deserves.

The conclusion of absence of symmetry breaking in the massive BRST formulation
is based on the observation that its typical formal indicators do not appear
in this setting \cite{Scharf}. The intrinsic (independent of the computational
method) argument for the presence of the Schwinger-Higgs screening is however
the vanishing of the Maxwell-charge associated with the identically conserved
(massive) Maxwell current. In contrast the conserved current in the Goldstone
spontaneous symmetry breaking model \textit{diverges} as the result of its
coupling to the massless Goldstone boson. The screening picture may not be
helpful in perturbative computations, but it quite easy to verify that the
mass gap in the BRST leads to the screened Maxwell charge.

The most exiting application of the stringlike formalism is certainly its use
for the clarification of the Higgs issue: are self-interactions of
vector-mesons only possible in the presence of a Higgs particle? The BRST
formalism leads to an affirmative answer \cite{Scharf} but it would be better
to understand this from the localization principle instead of leaving it to
the consistency of the indefinite matric BRST formalism. If this result is not
confirmed, all the hype about the world of QFT breaking down without a Higgs
particle was irrelevant and the presentation of LHC results with two competing
theoretical results would liberate particle physics from its self-generated
religious-ideological freight. If on the other hand the stringlike setting
confirms the necessity of the Higgs, it will present the best starting point
to find out whether the presence of \textit{low spin satellites} of
(self)interacting higher spin massive particles is a general consequence of
modular localization or whether this is a peculiarity of selfinteracting
massive vectormesons.

Apart from some mathematical problems as adjusting the Epstein-Glaser
renormalization framework to the needs of string-localization \cite{Mund}, the
calculations for most of the mentioned problems are still in progress
\cite{col}. Stringlike perturbation requires new concepts and is more
involved; the pointlike massive fields in terms which one hopes to express the
result of massive vectormeson interactions are not the the same fields as
those which are used to stay below the power-counting limit.

As one does not have to understand details about causal localization for
handling the renormalization theory of pointlike fields, one expects to find
rules for perturbing couplings involving stringlike fields which avoid the
intricacies of modular localization and lead to a simple description between
stringlike fields needed for renormalization and their pointlike counterparts.
Since in \cite{MSY} the idea of their use originated from the desire of a more
detailed understanding of the work on modular localization \cite{BGL} in the
setting of Wigner's representation theory, this connection with problems of
"real particle physics" alluded to in this section may also raise the readers
interest in the more abstract presentation of modular localization in the
following sections.

It is somewhat harder to speculate whether there could be some physical use
for the "kinematical" strings which are the generator of the string-localized
Wigner infinite spin representation. It is hard to imagine that any kind of
compactly localized counter can detect them or that they can be produced from
local interactions of ordinary particles. Weinberg \cite{Weinbook} dismissed
the infinite spin representations on the ground that "nature does not use
them", but it is questionable whether in times of dark matter one can uphold
this dismissal.

A yet different kind of spacelike strings arises in d=1+2 Wigner
representations with anomalous spin \cite{Mu1}. The modular localization
approach preempts the spin-statistics connection already in the one-particle
setting, namely if s is the spin of the particle (which in d=1+2 may take on
any real value) then one finds for the connection of the symplectic complement
with the causal complement the generalized duality relation
\begin{equation}
ZK(\mathcal{O}^{\prime})=K(\mathcal{O})^{\prime}%
\end{equation}
where the square of the twist operator $Z=e^{\pi is}~$is easily seen (by the
connection of Wigner representation theory with the two-point function) to
lead to the statistics phase $=Z^{2}$ \cite{Mu1}. Here the strings have a more
virtual existence, they serve to keep track of localization of plektons when
the representation of the universal covering group is activated by anomalous
spin but differential geometry does not provide a spacetime covering.
Different from all previous cases these representations have no functorially
associated free fields; surprisingly the QFT of anyons and plektons
(nonabelian representations of the braid group) do not even permit integrable
models \cite{B-M}.

\section{Algebraic aspects of modular theory}

A net of real subspaces $K(\mathcal{O})$ $\subset$ $H_{1}$ for a finite spin
(helicity) Wigner representation can be "second quantized"\footnote{The
terminology "second quantization" is a misdemeanor since one is dealing with a
rigorously defined functor within QT which has little in common with the
artful use of that parallellism to classical theory called "quantization".}
via the CCR (Weyl) respectively the CAR quantization functor; in this way one
obtains a covariant $\mathcal{O}$-indexed net of von Neumann algebras
$\mathcal{A(O)}$ acting on the bosonic or fermionic Fock space $H=Fock(H_{1})$
built over the one-particle Wigner space $H_{1}.$ For integer spin/helicity
values the modular localization in Wigner space implies the identification of
the \textit{symplectic} complement with the \textit{geometric} complement in
the sense of relativistic causality, i.e. $K(\mathcal{O})^{\prime
}=K(\mathcal{O}^{\prime})$ (spatial Haag duality in $H_{1}$). The Weyl functor
takes this spatial version of Haag duality into its algebraic counterpart,
whereupon the symplectic complement passes to the von Neumann commutant subalgebra.

One proceeds as follows: for each Wigner wave function $\varphi\in
K(\mathcal{O})\in H_{1}$ the associated (unitary) Weyl operator is defined as%
\begin{align}
&  Weyl(\varphi):=expi\{a^{\ast}(\varphi)+a(\varphi)\}\in B(H)\\
\mathcal{A(O})  &  :=alg\{Weyl(\varphi)|\varphi\in K(\mathcal{O}%
)\}^{^{\prime\prime}},~~\mathcal{A(O})^{\prime}=\mathcal{A(O}^{\prime
})\nonumber
\end{align}
where $a^{\ast}(\varphi)$ and $a(\varphi)$ are the usual Fock space creation
and annihilation operators of a Wigner particle with wave function $\varphi$.
This is a functorial relation between localization subspaces of the
one-particle space and localized subalgebras. Defining the algebra in terms of
the double commutant converts it into a von Neumann algebra i.e. a weakly
closed operator algebra.

This functorial relation between real subspaces and von Neumann algebras via
the Weyl functor preserves the causal localization and commutes with the
improvement of localization through intersections $\cap$ according to%
\[
K(\mathcal{O})=\cap_{W\supset O}K(W),~\mathcal{A(O})=\cap_{W\supset
O}\mathcal{A}(W)
\]
The functorial relation can be conveniently expressed in the commuting diagram%
\begin{align}
&  \left\{  K(W)\right\}  _{W}\longrightarrow\left\{  \mathcal{A}(W)\right\}
_{W}\label{cd}\\
&  \ \ \downarrow\cap~~~\ \ \ \ \ \ \ \ \ \ ~\ ~\downarrow\cap\nonumber\\
~~  &  \ \ \ K(\mathcal{O})\ \ \ \longrightarrow\ \ ~\mathcal{A(O})\nonumber
\end{align}
Here the vertical arrows denote the tightening of localization by
intersections whereas the horizontal ones stand for the action of the Weyl
functor. This commuting diagram expresses the functorial relation between
particles and fields in the absence of interactions and represents a
straightforward functorial extension of Wigner's representation
theory\footnote{Here we consider modular localization as part of Wigner's
theory because the modular localization is constructed within positive energy
representations of the Poincar\'{e} group \cite{BGL}.}. In the interacting
case this functorial connection is lost and the particle-field relations
becomes significantly more subtle. Its conceptual/mathematical complexity in
the case of non-integrable interactions is to blame for the 80 years lasting
lack of progress in proving even the mathematical existence of a model behind
its Lagrangian description, not to mention the problem of controlled
approximations. A new attempt concerning this age-old problem which is based
on the distinguished role of wedge-localized subalgebras will be presented in
the next section.

The case of half-integer spin representations is analogous \cite{Fa-Sc}, apart
from the fact that there is a mismatch between the causal and symplectic
complements which must be taken care of by a \textit{twist operator }$Z$ ; as
a result one has to use the CAR functor instead of the Weyl functor. In d=1+2
one encounters an exception; the Bargman-Wigner representation theory permits
anomalous spin which turns out to be connected with braid group statistics. As
already mentioned, this is the only known case for which (for $s\neq
$semiinteger)$~$there is no functorial relation between localized subspaces
and localized von Neumann subalgebras.

In case of the large family of irreducible zero mass "infinite spin"
representations, for which the lightlike "little group" is faithfully
represented, the functorial relation leads to string-localized generating
fields which generate the Fock space and the non-compact localized subalgebra
acting in it. There is an argument which excludes the existence of
compact-localized (generated by pointlike composites) subalgebras \cite{MSY},
but unfortunately it is not conclusive.

A discrete basis of local covariant field coordinatizations is defined by Wick
composites of the free fields. The case which deviates furthest from classical
behavior is the pure stringlike infinite spin representation for which the
class of relative string-localized fields form a \textit{continuous} family of
composites. Its non-classical aspects, in particular the absence of a
Lagrangian, is the reason why the spacetime description in terms of
semiinfinite string fields has been discovered only more than 60 years after
it appeared in Wigners classification \cite{BGL}\cite{MSY}.

Using the standard notation $\Gamma$ for the second quantization functor which
maps real localized (one-particle) subspaces into localized von Neumann
algebras, and extending this functor in a natural way to include the images of
the $K(\mathcal{O})$-associated modular objects (for which for which we use
the same notation $S,\Delta,J),$ one arrives at a special case of the Tomita
Takesaki modular theory for the interaction-free standard pair ($\mathcal{A(O}%
),\Omega$)\footnote{The functor second quantization functor $\Gamma$ preserves
the standardness i.e. maps the spatial one-particle standardness into its
algebraic counterpart.}%
\begin{align}
&  H_{Fock}=\Gamma(H_{1})=e^{H_{1}},~\left(  e^{h},e^{k}\right)
=e^{(h,k)}\label{mod}\\
&  \Delta=\Gamma(\delta),~J=\Gamma(j),~S=\Gamma(s)\nonumber\\
&  SA\Omega=A^{\ast}\Omega,~A\in\mathcal{A}(\mathcal{O}),~S=J\Delta^{\frac
{1}{2}}\nonumber
\end{align}

The Tomita-Takesaki theorem is about the action of the two modular objects
$\Delta^{it}$ and $J$ on the algebra%
\begin{align}
\sigma_{t}(\mathcal{A(O}))  &  \equiv\Delta^{it}\mathcal{A(O})\Delta
^{-it}=\mathcal{A(O})\\
J\mathcal{A(O})J  &  =\mathcal{A(O})^{\prime}=\mathcal{A(O}^{\prime})\nonumber
\end{align}
in words: the reflection $J$ maps an algebra (in standard position) into its
von Neumann commutant and the unitary group $\Delta^{it}$ defines an
one-parametric automorphism-group $\sigma_{t}$ of the algebra. In this form
(but without the last geometric statement involving the geometrical causal
complement $\mathcal{O}^{\prime})$ the theorem hold in complete mathematical
generality for "standard pairs" ($\mathcal{A},\Omega$). The free fields and
their Wick composites are "coordinatizing" singular generators of this
$\mathcal{O}$-indexed net of operator algebras in the sense that the smeared
fields $A(f)$ with $suppf\subset\mathcal{O}$ are (unbounded operators)
affiliated with $\mathcal{A(O})$ and generate $\mathcal{A(O})$ in an
appropriate mathematical sense$.$

Within the classifications of von Neumann algebras these local algebras are of
a very different type as their global counterpart. The latter is of the same
type as quantum mechanical algebras, namely an algebra of all bounded
operators on a Hilbert space $B(H)$. The local subalgebras are however
isomorphic to the aforementioned monad \cite{Haag}. More important than
understanding its position within Connes type classification for our purpose
is to characterize a monad by its use in physics. This is facilitated by the
fact that there are only two types which one needs for the formulation of
continuous\footnote{For discete combinatorial algebraic structures, as one
encounters them in lattice theories, also type II$_{1}~$enters, see remarks in
last section.} quantum physics: type $I_{\infty}$ (or $B(H)$, the algebra of
all bounded operators on a Hilbert space) in QM, and monads as localized
subalgebras in QFT \cite{Jakob}, where only the global algebra is a $B(H).$

The relevance of the T-T modular theory for interacting QFT is based on the
standardness of ($\mathcal{A(O}),\Omega$) (more generally for all finite
energy states) which is a consequence of the Reeh-Schlieder theorem
\cite{Haag}. The definition of the Tomita involution $S$ through its action on
the dense set of states (guarantied by the standardness of $\mathcal{A}$) as
$SA\Omega=A^{\ast}\Omega,$ and the action of the two modular objects
$\Delta,J$ (\ref{mod}) is part of the general setting of the modular
Tomita-Takesaki theory of abstract operator algebras in "standard position";
standardness is the mathematical terminology for the physicist's
Reeh-Schlieder property i.e. the property of that the action of a localized
subalgebra on the vacuum vector\footnote{In QFT any finite energy vector
(which of course includes the vacuum) has this property, as well as any
nondegenerated KMS state.} $\Omega\in H$ is cyclic $\overline{\mathcal{A(O}%
)\Omega}=H$ and $\mathcal{A(O})$ contains no "annihilators" of $\Omega.$

The important property which renders this useful as a new constructive tool in
the presence of interactions, is that for $\left(  \mathcal{A}(W),\Omega
\right)  ~$ the antiunitary involution $J$ depends on the interaction, whereas
$\Delta^{it}$ continues to be uniquely fixed by the representation of the
Poincar\'{e} group i.e. by the particle content. In fact it has been known
since Jost's seminal work on TCP \cite{Jost} (including the TCP covariance of
the S-matrix) that the interacting TCP operator is related to its free
(incoming) counterpart through the S-matrix; the $J,$ which represents the
reflection on the edge of the wedge, only differes from TCP by a $\pi
$-rotation. Rewritten in terms of the reflection $J$ on the edge of a wedge
this reads as%
\begin{equation}
J=J_{0}S_{scat} \label{scat}%
\end{equation}
In this form it attributes the role of a relative modular invariant (between
the interacting and free wedge-localized algebra) to the S-matrix and as a
result became a constructive tool of QFT \cite{Sch1}.

It is precisely this "semilocal" property of $S_{scat}$ in connection with
wedge-localization which opened the way for the inverse scattering
construction within the setting of the bootstrap-formfactor project.

The physically relevant facts emerging from modular theory in the general
setting can be condensed into the following statements:

\begin{itemize}
\item \textit{The domain of the unbounded operators }$S(\mathcal{O})$\textit{
is fixed in terms of intersections of the wedge localized algebras
}$\mathcal{A(O})=\cap_{W\supset\mathcal{O}}\mathcal{A}(W).$\textit{ The
domains associated to }$S(W)~$\textit{and }$domS(W)$\textit{ is}
\textit{determined by the representation of the Poincare group (and hence by
the particle content alone). These dense domains change with }$\mathcal{O}%
$\textit{ i.e. the dense set of localized states has a bundle structure.}

\item \textit{The complex domains }$DomS(\mathcal{O})=K(\mathcal{O}%
)+iK(\mathcal{O})$\textit{ in }$H_{Fock}~$\textit{decompose into real
subspaces }$K(\mathcal{O})=\overline{\mathcal{A(O})^{sa}\Omega}.$\textit{ This
decomposition contains dynamical information which in case }$\mathcal{O}%
=W$\textit{ includes the }$S_{scat}$\textit{-matrix (\ref{scat}). In the next
section arguments will be presented which suggests that with the help of a new
emulation formalism, which extends Wigner's representation approach to the
presence of interactions, the }$S_{scat}$\textit{-matrix under appropriate
conditions may fix }$\mathcal{A}(W)$\textit{ uniquely.}

\item \textit{ The restriction of the vacuum state to a local operator algebra
}$\mathcal{A(O})$\textit{ leads to a KMS relation at inverse modular
temperature }$\beta=1$%
\[
\left\langle AB\right\rangle =\left\langle Be^{-H_{mod}}A\right\rangle
,~e^{-itH_{mod}}:=\Delta^{it}%
\]
\textit{This localization-caused thermal behavior is accompanied by an area
proportional localization-entropy (section1). It has a variety of important
physical consequences.}
\end{itemize}

Modular localization is intimately connected to the holistic
aspect\footnote{For emphasising the importance of this property for the issue
of the cosmological constant we refer to the paper: "Quantum Field Theory Is
Not Merely Quantum Mechanics Applied to Low Energy Effective Degrees of
Freedom" by Hollands and Wald\ \cite{Ho-Wa}.} of QFT which places this theory
into a sharp contrast with QM (even in relativistic QM in the form of the
\textit{direct particle theory} (DPI) \cite{interface}. A one-dimensional
quantum mechanical chain or string of oscillators can be embedded into a space
of arbitrary dimensions since quantum mechanical (Born) localization is not an
intrinsic aspect of QT. On the other hand an embedding of a lower-dimensional
into a higher dimensional QFT is not possible; or to phrase it the other way
around: the restriction of a QFT to a lower dimensional submanifold
"remembers" (as a result of its holistic nature) that it is the restriction of
a more complete theory. In particular it is not possible to embed a
one-dimensional chiral theory into a higher dimensional QFT; a fact which is
overlooked in the incorrect picture of an embedding of a chiral theory into
its inner symmetry space ("target space") used in string theory (see the last
section for more remarks).

The holistic intrinsic nature of QFT presents itself most forcefully in the
possibility of characterizing a quantum field theory by the positioning of a
finite number of copies of an abstract monad in a shared Hilbert space. A
"modular inclusion" of one monad into another defines a chiral QFT, for a
3-dimensional theory one needs 4 modular-positioned monads, and placing 7
monads into a specific modular position leads to d=1+3 QFTs \cite{Kaehler}%
\cite{interface}. \textit{The positioning in Hilbert space determines not only
the algebraic substrate (the kind of quantum matter) of a QFT, but also
reveals its Minkowski spacetime localization properties and the action of the
Poincar\'{e} group on it}. The interpretation of a modular inclusion of two
monads is context-dependent; if one does not place additional monads into the
same Hilbert space, it defines a chiral theory on which the M\"{o}bius group
acts; a larger number of appropriately placed monads leads to higher
dimensional local nets of operator algebras. It is this intrinsic relation of
the abstract algebraic modular positioning of a finite number of monads in a
Hilbert space\ to the concrete localization of quantum matter in spacetime
(GPS positioning for LQP) that provides the strongest illustration of
"holistic"; there is no other theory of quantum matter which is "relational"
in this extreme way. As mentioned it forbids embedding of a lower into a
higher dimensional QFT and places severe restrictions on "dimensional
reduction" in QFT. Quantization is not a boundless game in which classical
manipulations can be interchanged with quantization; the holistic nature of
QFT shows its limitation and at the same time opens new perspectives. The
problem is that one cannot see these limitations on the level of Lagrangian
quantization; they would become visible if one tries to \textit{"curl up"
extra dimensions in explicitly computed correlation functions} by a
mathematically controlled operation on their correlations function which
maintains the holistic nature of QFT\footnote{If the model has sufficient
analyticity properties which allow real/imaginary time Wick-rotations, one can
"curl up" a time component by taking the high temperature limit in a KMS state
und create a new time direction by Wick rotation. But this is not what the
proponents of Kaluza-Klein reductions in QFT have in mind.} instead of
manipulationing Lagrangians, which is the way the Kaluza-Klein dimensional
reduction works in classical field theory.

It is interesting to briefly look at the difficulties which our QFT ancestors
encountered with these holistic aspects. From the time of the "Einstein-Jordan
conundrum" \cite{hol} through Jordan's subsequent discovery of QFT,
Heisenberg's discovery of vacuum polarization, Unruh's Gedankenexperiment and
Hawking's radiation up to the problem of the origin of the cosmological
constant, in all those cases the holistic nature of QFT asserts itself.

\section{"Emulation" as an extension of Wigner's representation theoretical
setting}

The \textit{bootstrap-formfactor program} \cite{Kar2} was based on the
assumption that the multiparticle components of a localized excitation of the
vacuum $A\left\vert 0\right\rangle ,$ $A\in\mathcal{A(O})$ in theories in
d=1+1 with an "integrable" S-matrix (purely elastic 2-particle
scattering\footnote{The elastic two-particle amplitude in d=1+1 is the only
scattering amplitude which cannot be distinguished from the identity
contribution by cluster factorization (equality of the product of two particle
plane wave inner products with the energy-momentum delta function), by which
factorizing models indicate their kinematical proximity to free fields.})
written in terms of rapidity variables are meromorphic functions which,
besides the degenerate represesentation of the permutation group from particle
statistics, possess a different nontrivial representation of this group under
\textit{analytic exchanges} (through analytic continuations) of rapidities.

This led the above authors to a change in notation in which the statistics
degeneracy is removed by encoding it into the left to right decreasing order
of the numerical values of rapidities
\begin{equation}
\left\langle 0\left\vert A\right\vert \theta_{1},\theta_{2},..\theta
_{k}\right\rangle ,~\theta_{1}>\theta_{2}>..>\theta_{k} \label{va}%
\end{equation}
so that other orderings can be used to represent the result of analytic
changes of ordering. The basic analytic change, the analytic transposition
between two adjacent $\theta,$ is defined in terms of a crossing symmetric
unitary two-particle S-matrix\footnote{These elastic S-matrices were obtained
from the classification of solutions of a "bootstrap" project for scattering
functions (solutions of the Yang-Baxter equations in case of matrix valued
scattering functions) \cite{Kar2}.}. From this analytic transposition one then
constructs an analytic representation of the permutation group. In contrast to
the degeneracy of the statistics representation which can be "dumped" into the
ordering prescription, this analytic representation is encoded into the
changes of orders of rapidities in the vacuum formfactor (\ref{va}).

Together with the crossing property which connects the multi-particle
components of local excitations of the vacuum $A\left\vert 0\right\rangle $
with particle formfactors between arbitrary multi-particle states together
with some general wisdom from LSZ scattering theory, this was basis of the
bootstrap-formfactor construction of a QFT. Within these rules the so called
inverse scattering problem admits a unique solution. In contrast to the
perturbative approach based on Lagrangian or functional quantization, the only
name for most of the models constructed in this way comes from their
scattering function; a few S-matrices (the most prominent is the Sine-Gordon
model) permit a perturbative relation with a Lagrangian from which these
models inherit their classical name.

This analytic representation of the permutation group called for an
algebraization in terms of a new kind of noncommuting particle operators. This
was formally achieved in \cite{Zam} and the associated algebra became known as
the Zamododchikov-Faddeev (Z-F) algebra (after Faddeev added a missing
c-number term). For a long time the physical interpretation of these operators
remained a mystery. Since the central property of QFT is \textit{causal
localizability,} it was natural to look for a relation of these generating
operators to the localization properties of QFT. The deviation of the Z-F
commutation relations from those of the standard creation/annihilation
operators exclude pointlike localization of their Fourier transforms.

As explained in the first two sections, a particular well-suited formulation
of QFT for unravelling the spacetime significance of operators is Haag's
"local quantum physics" (LQP) which places a net of localized observable
algebras and their "charged" representation sectors (from which together with
the observables one can construct a charge-carrying "field-net") into the
center stage and assigns to pointlike covariant fields the role of generators
of localized algebras without attaching a preferential status to a particular
field (apart from conserved currents which arise from the local implementation
of global symmetries). An important step in the development of an intrinsic
algebraic description was the aforementioned observation by Bisognano and
Wichmann \cite{Bi-Wi} (and its subsequent application to the Unruh effect and
Hawking's black thermal radiation by Sewell \cite{Sew}) that localization
within a spacetime wedge has a deep relation to the Tomita-Takesaki modular
theory of operator algebras including its thermal (KMS) aspects.

The application of these ideas to the bootstrap-formfactor project resulted
\cite{AOP} in the identification of the Fourier transforms of the Z-F
operators with generators of wedge algebras for integrable models. As free
fields, their one-time application to the vacuum state creates a one-particle
state, but different from the former, the presence of infinite vacuum
polarization clouds in interacting QFTs can in general not be avoided in their
iterative application to the vacuum. The Z-F operators turned out to be
special realizations of \textit{polarization-free-generators} (PFGs) in which
the relation between particles and fields resembled those in free field models.

A subsequent more foundational study based on "modular localization"
\cite{BBS} revealed that there exist two types of PFGs, temperate ones, whose
domains are translation invariant and whose particlelike Fourier transforms
fulfill Z-F algebra commutation relations), and nontemperate PFGs which exist
as operator-valued distributions only on wedge-localized test
functions\footnote{Unlike quantum fields they are not (operator-valued)
Schwartz distributions.} whose analytic on-shell restriction are only dense in
the $L^{2}~$integrable space of all wave functions. It turns out that
temperate PFGs only exist in d=1+1 and correspond to the family of integrable
models (factorizing models) of the bootstrap-formfactor program. Within this
restricted set of temperate wedge-generators, the so-called inverse problem
($S_{scat}\rightarrow QFT$) has a unique solution in that the (necessarily)
elastic $S_{scat}$ determines precisely one net of operator algebras (or a
Borchers equivalence class of local fields \cite{Haag}). The simpler
particle-field connection, which comes with temperateness and accounts for all
models which are integrable in the sense of the bootstrap-formfactor program,
led to view temperateness of PFG generators of wedge-localized algebras as
\textit{the~}foundational characterization of integrability of the associated QFT.

This new understanding of integrability in terms of localization reached its
present final touch in the work of Lechner \cite{Lech1}\cite{Lech2} who, by
solving the existence problem (the existence of a nontrivial net of compact
localized operators by intersecting wedge algebras) for many integrable
models, contributed an important constructive step to the almost 90 year
search for the mathematical control of QFT.

The resulting setting of integrable QFTs should be viewed as the
generalization of Wigner's representation theoretical approach for the
classification of one-particle spaces and a functorial construction of free
field algebraic nets in the presence of interactions (see previous section).
Wigner's motivation was that local quantum physics, as being more fundamental
than classical physics, should not be subjected to the parallelism to
classical theory implied in Lagrangian or functional quantization; in this he
echoed Jordan's 1929 dictum \cite{Jor} that QFT has to be formulated "without
classical crutches"; in the present context this is done in terms of a
concrete representation theoretical construction which is reminiscent of
Wigner's method in the absence of interactions.

The reformulation of the analytic bootstrap-formfactor project for integrable
models into a representation theoretical setting in terms of one-particle PFGs
as generators of wedge-localized operator algebras constitutes a successful
step in the adaptation of Wigner's idea to the realm of interactions. Although
nothing is known about the solution of the inverse problem within the vast
area of nontemperate (=nonintegrable) models (see later), one may ask whether
among all the QFTs which can possibly be connected with a given unitary
crossing symmetric S-matrix there is one for which the above idea of an
\textit{analytic change of order} permits a generalization.

To formulate such a problem one must first generalize the situation in
\cite{BBS} from one-particle PFGs to multi-particle states. This can be done
by using the same concepts from modular localization, since the dense set of
states of the modular Tomita operator $S$ also includes a dense sets of
multi-particle states of arbitrary high particle number which form a basis in
the Wigner-Fock Hilbert space of a QFTs with a complete particle
interpretation. Whereas in a later section we will also consider interacting
conformal QFTs which are known to have no particle interpretation and lead to
a different kind of only \textit{partial} integrability, in the present
context "integrability" (without an additional specification) will always
refer to the integrability in the sense of temperate particle wedge-localized
PFGs in QFTs \textit{describing particles}. In any interacting QFT with a
complete particle interpretation in terms of a incoming free field algebra
$\mathcal{A}_{in}$ there holds the following theorem

\begin{lemma}
(\cite{BBS}) Any state $\left\vert \psi\right\rangle \in domS_{\mathcal{A}%
(W)}=domS_{\mathcal{A}_{in}(W)}=dom\Delta^{\frac{1}{2}}~$can be generated from
the vacuum by operators $A~$and $A_{\mathcal{A}(W)}~$in each of the two
algebras%
\begin{align}
&  \left\vert \psi\right\rangle =A\left\vert 0\right\rangle =A_{\mathcal{A}%
(W)}\left\vert 0\right\rangle ,~A\in\mathcal{A}_{in}(W),~A_{\mathcal{A}(W)}%
\in\mathcal{A}(W)\label{lem}\\
&  A\overset{bijec}{\leftrightarrow}A_{\mathcal{A}(W)}%
~closed,~affiliated~to~\mathcal{A}_{\mathcal{A}(W)}\nonumber
\end{align}

\end{lemma}

Here the second line defines the "emulated" operators $A_{\mathcal{A}(W)}$ as
the image of wedge-localized free field operators under the bijection; the
Reeh-Schlieder property of the commutant algebras together with modular
operator theory (previous section) insures the denseness of their domains;
their closability (for their closure the same notation will be maintained) and
the unique affiliation with $\mathcal{A}(W)$ is clinched by the appropriately
formulated commutativity of the closed $A_{\mathcal{A}(W)}$ with the Haag-dual
commutant $\mathcal{A}(W)^{\prime}=\mathcal{A}(W^{\prime}).$ It is precisely
this properly defined \textit{commutativity with the commutant of the
interacting algebras }which\textit{ }insures\textit{ }the affiliation of the
emulats\textit{ }to the interacting algebra of the chosen model and not to any
other interacting algebras which happens to have the same particle content.
Modular localization of states is much weaker that localization of operators,
the two concepts only meet in the absence of interactions.

The crucial question is: how much does one have to know about the structure of
the interacting algebra $\mathcal{A}(W)$ in order to construct the emulats; is
the knowledge of the S-matrix in $J=S_{scat}J_{0}$ already enough for their
explicit construction? We know that for integrable models the particle
conserving two-particle scattering functions determine the Z-F algebra and
hence the QFT model, but can one also expect such a situation for
nonintegrable models when $S_{scat}$ describes particle creation? We will
present some consistency arguments which support such an idea, but are still
far away from a proof. These arguments involve quite novel ideas and so we
hope that the reader will find it sufficiently interesting to follow them.

Note that these unbounded emulats, whose unique existence in a given
interacting theory with a complete particle interpretation is guarantied, do
not by themselves form an algebra since they neither can be multiplied (even
in cases where this is possible $\left(  AB\right)  _{\mathcal{A}(W)}\neq
A_{\mathcal{A}(W)}B_{\mathcal{A}(W)}$), nor does the formation of emulats
commute with taking adjoints $(A^{\ast})_{\mathcal{A}(W)}\neq(A_{\mathcal{A}%
(W)})^{\ast}$\footnote{In fact it is easy to see that (\cite{BBS}) $\left(
A_{\mathcal{A}(W)}\right)  ^{\ast}\left\vert 0\right\rangle =S_{scat}%
A\left\vert 0\right\rangle $}. Although the original algebra cannot be
reconstructed directly, the objects obtained from polar and spectral
decompositions of the $\mathcal{A}(W)$ affiliated emulats can be used for its construction.

The hope that $\mathcal{A}(W)$ can be reconstructed solely from $S_{scat}$ is
based on the before mentioned idea of an analytic ordering change. An
important preparatory step is the formulation of the KMS propery in terms of
emulats and its relation to the crossing property of formfactors. For this it
is convenient to introduce the following notation, for simplicity we stay in
d=1+1. With%

\begin{align}
&  A(f_{1},.f_{n})\equiv:A(f_{1})...A(f_{n}):,~A(f_{1},.f_{n})\in
\mathcal{A}_{0}(W),~suppf_{i}\subset W\\
&  A(f_{1},.f_{n})\left\vert 0\right\rangle =\left\vert \check{f}%
_{1},.\right\rangle _{in},~~A(f)=\int a^{\ast}(\theta)\check{f}(\theta
)d\theta+h.c.,~p=m(ch\theta,sh\theta)\nonumber
\end{align}
denoting Wick-products of wedge-smeared free fields, these operators applied
to the vacuum create n-particle states in momentum space wave functions
$\check{f}$ which are the mass shell restrictions of the Fourier transform of
the $f$ in terms of rapidity variables $\theta.$ With $B$ denoting a generic
operator in the free field algebra $B\in\mathcal{A}_{0}(W)$\footnote{For free
fields the operator algebra is generated by exponential Weyl operators.}$~$or
an affiliated Wick-ordered composite, the KMS relation (section 4) in a (for
later purposes) useful form for the product of three operators reads%
\begin{equation}
\left\langle BA(f_{n},.f_{l+1})A(f_{l},.f_{1})\right\rangle \overset{KMS}%
{=}\left\langle A(f_{l},.f_{1})\Delta BA(f_{n},.f_{l+1})\right\rangle
,~B~composite\in\mathcal{A}_{0}(W) \label{r}%
\end{equation}
where the existence of the analytically continued boost $\Delta=e^{-2\pi
K_{W}},$ ($K_{W}=$ generator of W-preserving Lorentz boost) inside the right
hand correlation functions is guarantied by modular theory or by explicit
calculation within the free field Wick ordering formalism. Applying the
$A(f_{l},..f_{1})$ on the right hand side of (\ref{r}) to the bra vacuum, the
wave functions inside the bra vector are the complex conjugate of the
$\check{f},$ which in turn (using the analyticity resulting from the wedge
localization) are identical to the analytically continued original wave functions.

Finally by absorbing $\Delta^{\frac{1}{2}}$of the $\Delta$ into the analytic
continuation of the complex conjugate antiparticle wave functions (here the
localization is important) we arrive again at the original wave functions:
\begin{align}
&  \int\int\check{f}_{1}(p_{1})..\check{f}_{n}(p_{n})\left\{  \left\langle
0\left\vert B\right\vert p_{1},.p_{n}\right\rangle -\left\langle p_{1}%
,.p_{l}\left\vert \Delta^{\frac{1}{2}}B\right\vert p_{n},.p_{l+1}\right\rangle
\right\}  +contr.=0\label{f}\\
&  \left\langle 0\left\vert B\right\vert p_{1},.p_{n}\right\rangle
=\left\langle p_{1},.p_{l}\left\vert \Delta^{\frac{1}{2}}B\right\vert
p_{n},.p_{l+1}\right\rangle =:\left\langle -p_{1},.-p_{l}\left\vert
B\right\vert p_{n},.p_{l+1}\right\rangle ,~p_{i}\neq p_{k}\nonumber
\end{align}
where the contraction terms arise from Wick-contractions between the two
Wick-products on the left hand side of (\ref{r}). The free field distributions
inside the curled bracket are actually square integrable; so the first line
can be extended to all square integrable wave functions which then results in
the second line. It is easy to see that in the presence of antiparticles the
only change in the above relation is the assignment of the bra-momenta to
antiparticles, in short $p\rightarrow\bar{p}.$

The interacting analog of the free KMS relation for modular wedge localization is%

\begin{align}
&  \left\langle BA^{(1)}{}_{\mathcal{A}(W)}A^{(2)}{}_{\mathcal{A}%
(W)}\right\rangle \overset{KMS}{=}\left\langle A^{(2)}{}_{\mathcal{A}%
(W)}\Delta~BA^{(1)}{}_{\mathcal{A}(W)}\right\rangle \label{KMS}\\
&  \left\langle BA^{(1)}{}_{\mathcal{A}(W)}A^{(2)}\right\rangle =\left\langle
A^{(2)}{}_{\mathcal{A}(W)}\Delta~BA^{(1)}\right\rangle =\nonumber\\
&  =(S_{scat}A^{(2)\ast}\Omega,\Delta^{\frac{1}{2}}BA^{(1)}\Omega
),~since~(A^{(2)}{}_{\mathcal{A}(W)})^{\ast}\Omega=S_{scat}A^{(2)\ast}%
\Omega\nonumber
\end{align}
Rewriting this relation in terms of emulats for multiparticle states and using
the fact that the emulats acting on the vacuum are multiparticle states, one
obtains a pre-form of a particle crossing relation in which the particle
content of the emulat in the middle of the left hand side is still unknown%

\begin{align}
&  \left\langle 0\left\vert B(A^{(1)}(\check{f}_{1},.\check{f}_{l}%
)_{\mathcal{A}(W)}\right\vert \check{f}_{l+1},.\check{f}_{n}\right\rangle
=\int\check{f}_{1}(p_{1}),.\check{f}_{l}(p_{l})~_{out}\left\langle \bar{p}%
_{n},.\bar{p}_{l+1}\left\vert \Delta^{\frac{1}{2}}B\right\vert \check{f}%
_{1}..\check{f}_{l}\right\rangle _{in}~\label{c}\\
&  \int..\int\frac{d^{3}p_{1}}{2p_{10}}..\frac{d^{3}p_{n}}{2p_{10}}\check
{f}_{1},..\check{f}_{n}\{\left\langle 0\left\vert B(A^{(1)}(p_{1}%
,.p_{l})_{\mathcal{A}(W)}\right\vert p_{l+1},.p_{n}\right\rangle -\nonumber\\
~~  &  -\left\langle \bar{p}_{n},.\bar{p}_{l+1}\left\vert \Delta^{\frac{1}{2}%
}B\right\vert p_{1}..p_{l}\right\rangle
\}~\ \ ~=0\ \ \ \ \ \ \ \ \ \ \ \ \ \ ~\nonumber
\end{align}

In the sequel it will be shown how an extension of the idea of "analytic
ordering change" solves this problem and also suggests a formula for
formfactors of emulats as bilinear forms between multi-particle states. As in
the integrable case, we assume that the vacuum formfactor is locally analytic.
The presence of cuts resulting from inelastic multiparticle thresholds
prevents such formfactors to be boundary values of meromorphic functions as in
the integrable case, but these threshold cuts do not force them to be worse
than locally square integrable. In this case we may pass from wedge-localized
analytic wave function to n-particle wave functions with ordered supports and
obtain
\begin{align}
&  \left\langle 0\left\vert B(A^{(1)}(\theta_{1},..\theta_{l}))_{\mathcal{A}%
(W)}\right\vert \theta_{l+1},.\theta_{n}\right\rangle _{in}\equiv\left\langle
0\left\vert B\right\vert \theta_{1},...\theta_{n}\right\rangle _{in}%
\label{ana}\\
~~  &  for~~\theta_{1}>\theta_{2}....>\theta_{n}\nonumber
\end{align}
without having to know anything about the result of analytic order changes
within the $\theta_{1},..\theta_{l}~$cluster and relative to the remaining
$n$-$l$ particle cluster. In contradistinction to the integrable case,
analytic changes through the multi-particle threshold cuts will inevitably be
path-dependent, in particular there will be no analytic representation of the
permutation group\footnote{A similar case in x-space occurs if one extends a
d=1+2 Wightman setting to fields with braid.group statistics.}.

Our free field illustration (\ref{f}) suggests that the desired algebraic
structure should retract to the Wick-formalism in case $S_{scat}=1$ and to the
integrable analytic formalism for purely elastic 2-particle scattering. So we
are looking for a formula which describes the action of a PFG (a one-particle
emulat for simplicity) on an n-particle state in terms of a sum of terms in
which the $\theta$-dependent creation- and annihilation- components pass
through a particle cluster in order to arrive at its natural ordered position.
Assuming $...\theta_{k-1}>\theta_{k}>\theta>\theta_{k+1}>...$ we seek a
formula for the action of the distributional rapidity space creation component
$C(\theta)$ of the PFG $A(\check{f})_{\mathcal{A}(W)}$ of the type
\begin{align}
&  C(\theta)\left\vert \theta_{1},..,\theta_{k},\theta_{k+1},..\theta
_{n}\right\rangle =\label{action}\\
&  \sum_{l}\int d\vartheta_{1}..\int d\vartheta_{l}F_{\theta_{1}..\theta
_{k},\theta}(\vartheta_{1},..\vartheta_{l})\left\vert \vartheta_{1}%
,..\vartheta_{l},\theta,\theta_{k+1},..\theta_{n}\right\rangle \nonumber
\end{align}
where the $F$ only depends on the $\theta_{i}$ which are bigger than $\theta$
(i.e. which have been passed to achieve the natural order). If $\theta$ is
already larger than all the others $\theta_{i}$ in the state, the creation
part simply adds a particle. If $C(\theta)$ has to pass through a k-cluster to
arrive at its ordered position the result is required to be of the above form
whereas the annihilation component $C(\theta+i\pi)$ of $A(\check
{f})_{\mathcal{A}(W)}$ leads to a delta function $\delta(\theta-\theta_{k})$
multiplied with an integrand $F_{\theta_{1}..\theta_{k-1},\theta+i\pi
}(\vartheta_{1},..\vartheta_{l})\left\vert \vartheta_{1},..\vartheta
_{l},\theta_{k+1},..\theta_{n}\right\rangle .$ The sum over l takes into
consideration that processes of passing through particle clusters create and
annihilate particle states of arbitrary high particle number, whereas the
$\theta_{k+1}..\theta_{n}$ on the right of the k-cluster remain unchanged.

The second requirement is that $F$ should not contain more detailed
information about the interacting algebra $\mathcal{A}(W)$ than those
contained in $S_{scat}$ which enters the wedge localization as a relative
modular invariant. If the emulats depend on more detailed properties of
$\mathcal{A}(W)$, they remains outside the range of the present method.

The still undetermined $F$ will now be specified in terms of the "grazing
shot" amplitude%

\begin{align}
F_{\theta_{1}..\theta_{k},\theta}(\vartheta_{1},..\vartheta_{l})  &  =\sum
_{s}\int d\chi_{1}..\int d\chi_{s}S^{\ast}(\chi_{1},..\chi_{s}\rightarrow
\vartheta_{1},..\vartheta_{l})\cdot\\
\cdot S(\theta_{1},..\theta_{k},\theta &  \rightarrow\chi_{1},..\chi
_{s},\theta)\nonumber
\end{align}
which consists of a product of a scattering amplitude, in which one particle
with rapidity $\theta~$scatters together with $k$ other particles such that
remains unchanged (second line). This is multiplied with the complex conjugate
of a second amplitude whose purpose is to compensate all processes which would
have occured even in the absence of grazing shot rapidity $\theta.$ Hence
without the presence of $\theta$ nothing happens i.e. $F$ reduced to (particle
matrixelements of) the unit operator.\ This construction allows an extension
from one-particle PFGs to multi-particle emulats.

In case of integrable models the S-matrix does not change the particle number
and the grazing shot S-matrix $F$ (\ref{action}) reduces to a multiplication
with a product of elastic S-matrices \cite{Kar2},
\begin{equation}
S(\theta-\theta_{1})..S(\theta-\theta_{k})
\end{equation}
where in the case of the action of the annihilation components one S-factor is
replaced by a delta contraction. In fact the whole idea behind the
construction is to obtain a formula for a general S-matrix which in the
integrable case passes to the known expression.

In the nonintegrable case this presentation is extremely formal since we know
from \cite{BBS} that nontemperate PFGs are only meaningful as wedge-localized
$A(f)_{\mathcal{A}(W)}$ acting on wedge-localized states. Hence it is better
to think of the PFGs as bilinear forms between particle states%
\begin{equation}
\left\langle \theta_{1}^{\prime},..\theta_{m}^{\prime}\right\vert
C(\theta)\left\vert \theta_{1},..\theta_{n}\right\rangle
\end{equation}
and restore the localizing wave functions which one needs for passing to
operators $A(f)_{\mathcal{A}(W)}$ after having computed these formfactors of
the $C(\theta)^{\prime}s.$

The rapidities are uniformization variables which remove the elastic
threshold, so that formfactors of integrable models are meromorphic functions
in the multi-$\theta$ plane. The general analytic ordering picture assumes
that the only singularities are cuts from higher inelastic thresholds which
are locally square integrable. But if locally square integrable rapidity wave
functions are admitted, we may extend the validity of the KMS relation from
wedge localized wave functions to wave functions with ordered square
integrable supports. In this way the analytic ordering picture connects the
formfactor crossing identity (including its analytic property) to the KMS
crossing identity for formfactors (\ref{ana})%

\begin{align}
\left\langle 0\left\vert B\right\vert p_{1}...p_{n}\right\rangle _{in}  &
=~_{out}\left\langle -\bar{p}_{l+1},.-\bar{p}_{n}\left\vert B\right\vert
p_{1},.p_{l}\right\rangle _{in}\\
for\text{ }(\theta_{1},..\theta_{1})  &  >(\theta_{l+1},..\theta_{n})
\end{align}

It is remarkable that the interacting crossing identity just looks like its
free field counterpart, apart from the fact the the bra-vectors refer to
outgoing particle. Its derivation is not affected by the fragile conceptual
status of our grazing shot construction for the action of emulats on
multi-particle states.

Its origin from the KMS property of wedge-localization shows that this
property in the center of particle theory shares its conceptual roots with
those of the thermal manifestation of localization (the Einstein-Jordan
conundrum of subvolume fluctuations, the Unruh Gedankenexperiment and Hawking
radiation from black hole event horizons). In particular it has no relation to
the crossing in the dual model and string theory which results from crossing
properties of (Mellin transforms of) conformal correlation functions (section
4). String theory is the result of a fundamental misunderstanding of the
subtleties of causal localization.

Although the grazing shot Ansatz reduces to the Wick contraction formula in
the absence of interactions and to relations obtained from the Z-F algebra in
the integrable case, these are only rather weak consistency requirements. The
crucial step in establishing its correctness is the verification of
wedge-locality%
\[
\left[  JA(f)_{\mathcal{A}(W)}J,A(g)_{\mathcal{A}(W)}\right]
=0,~suppf,g\subset W
\]
which, apart from the trivial case of vacuum and one-particle matrix elements,
the author was unable to do. This property may also turn out be useful for a
perturbative determination of of PFGs $A(W)_{\mathcal{A}(W)}$ which one
expects to be analog the iterative use of causality in the Epstein-Glaser
iteration for pointlike fields. If the divergence of the on-shell perturbation
series would be related to the pointlike singular character of fields, one
expects that on-shell perturbation theory for wedge-local operators should converge.

Hence the grazing shot construction still hangs in the air and it is presently
not worthwhile to present this idea in more details. In fact the reason why
this is mentioned here at all is that the author firmly believes that even
incomplete or failed attempts on important problems in particle physics should
not go unmentioned; reporting on them is not less important than presenting
established facts. The proof of mathematical existence of nontrivial models
outside the narrow setting of integrability and the discovery of controlled
approximations remains still the paramount problem of QFT even after almost 90
years after its inception. The resounding \textit{observational success}
resulting from low orders of the (unfortunately) \textit{diverging}
perturbative series has in no way disburdened QFT from its
conceptual-mathematical fragility; the idea that the low orders are an
asymptotic solution in the limit of vanishing interaction strength remains an
unproven conjecture, even after Dyson pointed to this problem more than half a
century ago.

In the integrable case the inverse problem has a unique solution, but there is
no argument that excludes the possibility that those elastic S-matrices also
admit \textit{non-integrable solutions}. The lack of a uniqueness proof even
includes the absence of interactions in the sense of $S_{scat}=1,$ i.e. the
question whether besides the massive free field, the Hilbert space can
accommodate other local nets with the same representation of the Poincar\'{e}
group and the TCP operator (and hence the same modular data\footnote{In
\cite{1977} it was shown that as a consequence of Huygens principle, conformal
QFT leads to the uniqueness of the inverse problem for $S_{scat}=1.$}.

Without additional restrictions it is in fact quite easy to construct a
continuous infinity of covariant wedge-localized algebras. Any unitary
operator in $H$ which preserves the vacuum and commutes with the
wedge-preserving Lorentz boost and the reflection $J_{0}$ on the edge of the
wedge of the form%
\begin{equation}
V=e^{i\eta},~\eta=\sum\frac{1}{n!}\int\tilde{\eta}(x_{1},..x_{n}%
):A(x_{1})..A(x_{n}):dx_{1}..dx_{n}%
\end{equation}
will lead to a net of wedge algebras with $S_{scat}=1\ $which, apart from the
special case that $V$ implements an automorphism of $\mathcal{A}_{0}(W),$ is
inequivalent to the net generated by a free field. The restrictions on the
coefficient functions in the rapidity parametrization $\tilde{\eta}(\theta
_{1},..\theta_{n})$ for d=1+1 resulting from the shared modular data are very
mild: the coefficient functions $\tilde{\eta}~$can only depend on $\theta
$-differences, the $J_{0}$-invariance leads to a reality condition and
$V\left\vert 0\right\rangle =\left\vert 0\right\rangle $ requires the absence
of terms with only creation operators.

The passing from a fixed $W$ to the net of wedges $W_{x}$ ($x=apex$) is
achieved by applying translations $U(x)$%
\begin{equation}
\mathcal{A}(W_{x})\equiv V_{x}\mathcal{A}_{0}(W_{x})V_{x}^{\ast}%
,~V_{x}=U(x)VU(x)^{\ast}%
\end{equation}
Unlike the previous method which was based on the temperate/nontemperate
dichotomy of emulats, it would be very difficult to separate integrable models
with $S_{scat}=1$ from this huge set of possibilities\footnote{The consistency
of this method with the emulation construction may lead to further
restrictions.}.

This argument only concerns the connection between $S_{scat}=1$ and the family
of wedge-localized local nets. In order to arrive at the full net, one still
has to intersect wedge algebras in order to obtain compact localized double
cone algebras $\mathcal{A(D)}$ and one knows from the work of Lechner
\cite{Lech2} that the requirements about the \textit{cardinality of phase
space degrees of freedom,} which insures the nontriviality of intersections,
are extremely restrictive; in fact most wedge nets will not possess nontrivial
double cone nets. This still leaves the possibility that the looked for
uniqueness may arise in the problem of forming intersections.

The aim of the present approach based on wedge-localization is closely related
to the old S-matrix setting; it shares with the ideas behind the abandoned
S-matrix bootstrap and Mandelstam's later attempts (to use spectral
representations for the description of analytic properties of elastic
scattering amplitudes\footnote{It is however incompatible with dual model and
string theory constructions and their S-matrix interpretations.}) its
proximity to laboratory observables. The S-matrix (formfactor of the identity)
and formfactors are on-shell objects (correlations restricted to the
mass-shell) which are directly accessible to experiments. They contain much
more information than Lagrangians which lead to a unique perturbative series.
The constructions in this section are thought of as a first step to construct
the full (off-shell) QFT.

This makes them \textit{top-to-bottom approaches} in the sense that one starts
with a list of well-understood properties, which one expects a QFT to
describe, and then sets up the mathematics to understand their consequences.
Pointlike localized fields and their correlations are far removed from
particles and their on-shell manifestations; they form the
\textit{bottom-to-top} setting of Lagrangian or functional quantization; in a
mass-shell based top-to-bottom approach they are only expected to appear at
the end (in case they are still needed). In quantization approaches one
follows a translation dictionary which forces a more fundamental QFT to follow
formal rules which correspond to those of a less fundamental classical theory.
As a result, the physical interpretation and content only emerges at the end
of perturbative calculations and there is hardly any mathematical controll
about what one is doing. A top to bottom approach as the present attempt cuts
all classical connections and replaces them by a tighter
conceptual-mathematical control.

Perhaps the conceptually most useful analogy to what is done here is to
consider it as an extension of the Wigner representation theoretical approach
to the presence of interactions. The functorial connection between Wigner
particle and free fields breaks down in the presence of any
interaction\footnote{This is different in QM where the interacting
Schr\"{o}dinger equation remains functorially related to its operator Fock
space formulation.}. The relation between particles and their free fields to
interacting wedge-localized emulats can be viewed as its substitute. This much
weaker particle-field connection just exposes an intrinsic difficulties of
interactions which prevented the construction of d=1+3 interacting fields. It
is certainly premature to think of attaching a catch name of this still very
tentative attempt of getting a nonperturbative grip on non-integrable
QFTs\footnote{A warning exemple is string theory whose name has nothing to do
with its content.}.

\section{Conformal integrability}

There exists a different notion of "kinematic" integrability which is not
directly related to the dynamics of a model, but rather refers to a discrete
combinatorial structure of its countable superselected localizable charge
sectors which the DHR superselection theory \cite{Haag} uniquely associates to
a local (neutral and invariant under inner symmetries) observable algebra.
More explicitly, it refers to the structure of the set of equivalence classes
of localizable representations of the observable net \{$\mathcal{A(O}%
)\}_{\mathcal{O}\in R^{4}}$ given in its vacuum representation. In the case of
massive theories with Bose/Fermi statistics this structure turns out to have
the form of a tracial state on a discrete (type II$_{1}$) operator algebra
which contains the infinite permutation group algebra whose representation
theory is responsible for the particle/field statistics. In fact this algebra
turns out to be the dual of a compact "internal symmetry" algebra which
commutes with the Poincar\'{e} group \cite{Haag} \cite{DR}. The internal
symmetry group acts on a larger uniquely determined "field algebra" which
contains the observable algebra as a fixed point algebra under the action of
the symmetry group \cite{DR}. This construction demystifies the concept of
inner symmetries, which dates back to Heisenberg's phenomenological
introduction of isospin into nuclear physics. Its conceptual origin in QT is
the principle of causal localization of quantum observables and their
superselected localized representations which can be combined into a field
algebra on which a compact group acts\footnote{To put it bluntly: compact
groups arise from quantum causal localization in the presence of mass gaps.
.}. Although it has no counterpart in classical physics, for the application
of the method of Lagrangian or functional quantization it is necessary to read
this property back into the classical setting.

In previous sections we have seen that causal localizability results in vacuum
polarization, thermal manifestations and associated intrinsic ensemble
probabilities. The DHR theory adds the superselection theory and inner group
symmetries of field algebras in which the observable algebras are embedded as
fix-point algebras under the action of the internal symmetry to this list.

The relation between local observables and their extension into a field
algebra becomes particularly interesting in theories which cannot be described
in terms of Lagrangian quantization; the most prominent family of such
theories are conformally invariant QFTs\footnote{The indirect way of
interpreting conformal theories as massless limits of perturbatively
accessible massive models was only successfull in the case of the massive
Thirring model for which the perturbative Callen.Symanzik equation comes with
a vanishing beta function \cite{Lo-Go}..}. In that case the observable
algebras obey the Huygens principle i.e. the vanishing of (graded) commutators
of pointlike fields also for timelike distances. Apart from chiral conformal
theories which live on lightlike lines so that the distinction between space-
and time- like disappears and for which the observable algebras (current or
energy-momentum algebras) as well as the generators of their representations
(braid group commutation relations) in many typical cases can be explicitly
constructed, the "Huygens" observable algebras in higher dimensions are more
complicated and do not seem to be integrable.

However the anomalous spectrum of scale dimension in conformal theories seems
to be susceptible to systematic classification. This is because it is defined
in terms of the phases of a unitary operator, the generator of the center $Z$
of the conformal covering group. Whereas the "Huygens observable algebra"
lives on the compactified Minkowski spacetime $M^{c}$, the field algebra,
which is generated by pointlike fields with anomalous dimensions, is localized
on its universal covering $\widetilde{M^{c}}$\footnote{Equivalently they can
be interpreted as operator-valued distributional sections on $M^{c}.$}$.$ As
in the simpler case of chiral models, where the nontrivial center is closely
related to the issue of plektonic statistics (braid-group commutation
relations of fields), one is accustomed to view such properties as being more
on the kinematic than on the dynamic side, although such distinctions become
somewhat blurred outside that kind QFT which is used for the description of
particles to which conformal QFT definitely does not belong. In the sequel we
will argue that the the spectrum of anomalous dimensions is indeed accessible
to rigorous classification and that therefore the terminology kinematical (or
better "partial") integrability of conformal QFTs is quite appropriate.
Integrability without additional specification will be reserved for the full
dynamical integrability which is limited to d=1+1 as explained in the previous section.

The method to investigate partial integrability in conformal QFT is
nonperturbative and, as the constructive approach to integrable massive
models, uses representation theory in particular the representation
theoretical methods of the DHR\ superselection theory. Already for the
low-dimensional chiral models the role of the infinite permutation group
$\mathbf{P}_{\infty}$ is taken over by the much richer representation theory
of the braid group $\mathbf{B}_{\infty}.$ As the DHR theory led to tracial
states ("Markov traces") on the infinite permutation group, chiral conformal
theories require a classification of representations associated with tracial
states on $\mathbf{B}_{\infty}.$

Tracial states on combinatorial algebras in a much more general context are an
important tool in Vaughn Jones subfactor theory \cite{Jones}. The partial
integrability of the braid group representation structure, even in cases where
it was not possible to compute n-point correlation functions of observable
fields\footnote{A notable exception is the chiral Ising QFT \cite{Re-Sch}.}
led to a mutually interesting and fruitful connection with subfactor theory
(which may be viewed as a vast extension of group representation theory).

Whereas $d\geq1+2$ interacting models with a complete particle interpretation
are always non-integrable (previous section), their superselection structure
which forms a discrete tracial algebra, is by definition integrable since
tracial states on words in extended group algebra of the permutation
$\mathbf{P}_{\infty}$ or braid group $\mathbf{B}_{\infty}$ are computable by
combinatorial methods. This is of particular interest in higher dimensional
conformal theories for which DHR superselection setting suggests the relevance
of a braid-permutation group $\mathbf{PB}_{\infty}$ whose
representation-theoretical studies are still in their its infancy \cite{Fenn}.

To investigate this partial integrability one needs to know some structural
properties of conformal QFTs, in particular that anomalous dimensions are
labels of superselection sectors of observable algebras. The braid group and
the spacetime covering aspect arises through the time-like Huygens structure,
whereas the permutation group enters as usual through spacelike commutativity;
but the $\mathbf{PB}_{\infty}$ group is not simply a product of its two
subgroups! A detailed knowledge about local observables is not required; it is
not necessary to know what is "inside" each superselection sector, the rules
of their compositions and decompositions suffice. The combinatorial algebras
of the Hecke or Birman-Wenzl type are typical algebraic structures which arise
in this context \cite{R-S}.

There have been some misconceptions in the recent literature about the status
of conformal QFT within particle theory\footnote{In many contemporary articles
the fact that the tree-approximation of conformal theory (isomorphic to the
classical structure) allow a restriction to a zero mass shell has been used to
incorrectly allege that they can describe quantrum particles in the sense of
scattering theory and the S-matrix.}. Conformal QFT was first proposed in the
beginning of the 60s, but as a result of their remoteness from particle theory
in particular scattering theory the interest in them waned quickly. Most of
the intuitive arguments against their direct use in particle theory were later
made rigorous. Here are some of them

\begin{enumerate}
\item A conformal field with canonical (free field) short distance behavior is
inevitably identical to a free field theory \cite{old}.

\item A conformal QFT cannot be perturbatively constructed from free massless
fields and the perturbative behavior of massive renormalizable d=1+3 models
(contrary to some models in d=1+1\footnote{The most prominent exeption is the
massive Thirring model. In fact the suspicion that $\beta(g)\equiv0$ which led
the derivation of the Callan-Symanzik equation to all orders \cite{Lo-Go} came
from the observation of softness in m$\rightarrow0.$}) is not "soft" in a
sense which would allow to take a massless limit within the perturbative
Lagrangian setting.

\item The LSZ scattering limits of interacting conformal fields
vanish\footnote{The Hilbert space positivity forces the K\"{a}ll\'{e}n-Lehmann
spectral measure to have a singularity which is milder than a mass-shell delta
function.}.
\end{enumerate}

The proof for 1. and 3. is actually quite simple, the first follows from the
fact that the canonicity of scale dimension requires a free field behavior for
short distances whin in conformal theories implies the freeness of the field
itself. The third is a consequence of the fact that the increase of the short
distance dimension above its smallest possible value allowed by positivity
(that for a free field) automatically reduces the singularity at the zero mass
shell $p^{2}=0$ which then is too weak to match the dissipating behavior of
wave packets which would be necessary in order to arrive at a nontrivial LSZ
limit. The QED prescription, which interprets photon-inclusive cross sections
as the observable manifestations of charged particle does not work for
assigning a particle interpretation to conformal theories.

The airy use of conformal QFTs in many recent publication shows that particle
physics is in the process of loosing its history since none of the old
arguments showing the problematic relation of conformal QFT with particle
theory has been addressed. It is not forbidden to think about conformal
theories of resulting from massive theories in a hypothetical limit in which
all particle creation threshold fall on top of each other, but as long as
there is no such massive theory with a perturbative Callan-Symanzik equation
with $\beta=0~($see remarks in section 3) this is of not much use.

Although conformal theories play no direct role in particle theory, their
apparent mathematical simplicity make them ideal "theoretical laboratories"
for the study of structural problems of QFT. Conformal transformations relate
compact to non-compact regions and in this way extends the concept of modular
localization. Low dimensional chiral theories were the first theories for
which representation theoretical nonperturbative methods led to proof of
existence as part of their explicit construction before such methods were also
successfully applied for massive integrable models (previous section). They
played an important role in the adaptation of the Tomita-Takesaki modular
theory of operator algebras to problems of modular localization, and led to a
fruitful meeting of minds between algebraic QFT and the subfactor theory.

An important step in the history of conformal QFT was the understanding of the
role of the Huygens principle in the definition of conformal observables and
anomalous dimension-carrying charged fields which led in 1975 to a conformal
decomposition theory \cite{S-S}\cite{S-S-V}\cite{Lu-Ma}. There were two
viewpoints about conformal invariance; one can either say that conformal
fields "live" (are univalued) on \textit{the universal covering of the
compactified Minkowski spacetime }$\widetilde{M^{c}},$ or that they are
distribution-valued sections on $M^{c}.$ In the first case \cite{Lu-Ma} (which
probably goes back to Irving Segal) one encounters infinitely many "heavens"
above and "hells" below $M^{c}~$and there exists a \textit{generator of the
center} of the universal conformal covering group $Z\in\widetilde{SO(4,2)}%
$\ (for d=1+3) such that $Z^{n},$ $n~integer~$numbers those heavens and hells
and $n=0$ corresponding to the compactification $M^{c}$ of our living
spacetime. The center is a certain \textit{conformal rotation} at the angle
$2\pi$ whose spectrum results in the formula $specZ=\left\{  e^{i2\pi
d_{\alpha}}\right\}  $ where $d_{\alpha}$ runs over the (anomalous)~conformal
field dimensions.

There is an analogy of this situation to the physics of plektons in d=1+2. In
this case the Poincar\'{e} group $\mathcal{P}$ has an infinite covering
$\widetilde{\mathcal{P}},$ but the spacetime has none. The Wigner-Bargmann
representation theory of positive energy representations in d=1+2 however
creates a kind of covering due to the semiinfinite string-like nature of the
plektonic wave functions \cite{Mund3}. Apart from that difference the
anomalous spatial spin corresponds to the anomalous dimension and the
plektonic statistics (anyons are abelian plektons) resembles an imagined
"timelike braided exchange", with the resulting statistical phase \cite{Haag}
corresponding to the eigenvalue of $Z.$ For a long time it was suspected that
there is some kind of free or at least integrable plektonic QFT associated
with the Wigner-Bargman representation, but meanwhile this idea has been
disproven \cite{B-M}.

The prerequisite for conformal observables is that their scale dimension is
integer\footnote{For semiinteger dimension as they already occur for free
spinors it is necessary to take the double covering of $M_{c}.$ These fields
fulfill an extended Huygens principle on the double covering.}; typically
their pointlike generators are conserved currents or the energy-momentum
tensor which result from the "localization" of global symmetries; but in
principle any field with integer dimension satisfies the Huygens propertyn and
hence can be included into the observables. Such local fields live on $M^{c}$
and commute with the center $Z~$of the conformal group $\widetilde{SO(4,2)}.$
As a result their commutators are concentrated on the mantle of the light cone
which in turn implies that their correlations functions are multivariable
rational analytic functions \cite{Ni-Re-To}. Despite their simple appearance,
nontrivial d=1+3 Huygens fields have not yet been constructed and the problem
of their integrability remains unresolved.

The anomalous dimensions play the role of generalized superselected charges
carried by the anomalous dimensional fields\footnote{The analogy works better
with squares of charges since the matter-antimatter charge compensation has no
counterpart the composition of anomalous dimensions.}. The application of the
spectral decomposition theory with respect to the center $Z$ leads to the
following decomposition of fields
\begin{align}
A(x)  &  \rightarrow A_{\alpha,\beta}(x)\equiv P_{\alpha}A(x)P_{\beta}%
,~Z=\sum_{\alpha}e^{i2\pi d_{\alpha}}P_{\alpha}\\
&  A_{\alpha,\beta}(x)B_{\beta,\gamma}(y)=\sum_{\beta^{\prime}}R_{\beta
,\beta^{\prime}}^{(\alpha,\gamma)}(x,y)B_{\alpha,\beta^{\prime}}%
(y)A_{\beta^{\prime},\gamma}(x)~\nonumber
\end{align}
The R-matrices depend discontinuously on spacetime, they are locally constant
but different for time- and space- like separations. For time-like separation
the distinction positive/negative timelike is topologically similar to
left/right distinction in chiral theories.

These decompositions appear first for abelian (anyonic) R-matrices in
\cite{S-S}; only after the path-breaking work in the 80s by Belavin,Polyakov
and Zamolodchikov \cite{BPZ} they were generalized to the nonabelian braid
group (plektonic) representations which appear in the exchange algebras of
chiral models \cite{R-S}\cite{F-G}. Although (apart from free fields) chiral
conformal field theories do not describe particles, it is customary to refer
to the quanta, which carry a discrete representation of the conformal rotation
and lead to R-matrix commutation relations of the Artin braid group, as "plektons".

The topological similarity of the positive and negative time-like Huygens
region leads one to expect the anomalous dimension spectra to be connected
with braid group representations. But since there is also the requirement from
spacelike commutation which leads to the nontrivially combined $\mathbf{BP}%
_{\infty}$ group (the "words" in $\mathbf{B}_{\infty}$ intertwine nontrivially
with those in $\mathbf{P}_{\infty}$), the kind of $\mathbf{B}_{\infty}$
representations as they occur in chiral theories (Hecke-, Birman-Wenzl
algebras,..) are not expected to re-appear in this form in higher dimensional
CFT. Some of the new problems were mentioned in \cite{old}.

It is not difficult to write the defining relation between the $b_{i},t_{i}$
$~$generators of $\mathbf{BP}_{\infty}~$\cite{Fenn}$.$ The $Z$ spectrum of any
4-dim. conformal model should belong to one representation of these relations.
But unlike in the chiral case, where the exponential Bose field model was
available a long time before the later systematic construction of families of
chiral models in the work of \cite{BPZ}, there exists presently no
illustrative nontrivial example; the conformal invariant \textit{generalized
free field} (which results from the AdS free field by applying the AdS-CFT
correspondence) is too far away from physical fields\footnote{Its abundance of
degrees of freedom leads to the before-mentioned pathological timelike
causality properties and the absence of reasonable thermodynamik behavior.} in
order to be of much interest. Recently proposed dimensional spectra on the
basis of analogies with those of transfer matrices of Ising lattice models
\cite{Bei} are not supported by the Huygens structure of conformal observables
and their expected dimensional spectra from the $\mathbf{BP}_{\infty}$
representation structure.

The group theoretical origin of the simplification in d=1+1 is the
factorization of its conformal group $SO(2,2)=SL(2,R)\times SL(2,R)~$which
leaves the 3-parametric Moebius group as the space-time symmetry of a chiral
theory on $\mathbb{R}$ or its compactification $S^{1}~$(with the possibility
to extend it to Diff(S$^{1}$)). The group theoretical factorization is
followed by a decomposition of the d=1+1 conformal observables into its chiral
components living on separate light rays. The chiral theories have proven to
be the most susceptible to the classification and construction of their their
superselected representation sectors and sector creating plektonic fields; in
particular for "rational" models (i.e. models with a finite number of
representations) many explicit constructions are available \cite{Ka-Lo}.

The first illustrative model for the decomposition theory was the exponential
Boson field \cite{S-S}. In this case the analogy of anomalous dimension with
superselecting charges takes a very concrete form. In a somewhat formal way of
writing
\begin{align}
j(x)  &  =\partial_{x}V(x),~~\left\langle j(x),j(x^{\prime})\right\rangle
\sim\frac{1}{(x-x^{\prime}+i\varepsilon)^{2}}\\
\Psi^{(q)}(x)  &  =e^{iqV(x)},~[Q,\Psi^{(q)}(x)]=q\Psi^{(q)}(x),~Q=\int
j(x^{\prime})dx^{\prime}\nonumber\\
\Psi^{(q)}(x)  &  =\sum_{q^{\prime}-q^{\prime\prime}=q}\Psi_{q^{\prime
},q^{\prime\prime}}^{(q)}(x),~\Psi_{q}^{q^{\prime}}(x)\equiv P_{q^{\prime}%
}\Psi_{q}(x)P_{q^{\prime\prime}}\nonumber
\end{align}
In the last line the $P_{q^{\prime}}$ are the projectors onto the continuous
subspaces $H_{q^{\prime}}$ where $q^{\prime}~$runs over all continuous
superselected charge values in a nonseparable Hilbert space. This model
belongs to the class of non-rational chiral models, but by enlarging the
observable algebra to include $\Psi^{(q)}(x)$ fields with $q^{\prime}%
s~$leading to integer scale dimensions $q^{2}$%
\begin{equation}
U(\lambda)\Psi^{(q)}(x)U(\lambda)^{\ast}=\lambda^{q^{2}}\Psi^{(q)}(\lambda x)
\end{equation}
the number of superselected sectors becomes finite (quantized charges) and the
resulting model is "rational" and integrable.

The situation becomes especially interesting for an n-component current
algebra.%
\[
j_{k}(x)=\partial_{x}V_{k}(x),~k=1..n,~\Psi^{(\vec{q})}(x)=e^{i\vec{q}\vec
{V}(x)}%
\]
In that case the maximal local extensions of the current algebra are
classified in terms of integer n-dimensional lattices and their superselected
sectors are characterized in terms of their dual lattices. In this setting the
selfdual lattices correspond precisely to situations with a trivial
superselecting structure (the vacuum sector is the only sector). The
corresponding selfdual lattices correspond to the largest exceptional finite
groups, the most mysterious among them is known as the "moonshine" group. The
fact that quantum localization lead to such subtle group theoretic properties
gives an impression of the conceptual depth and its many unexpected relation
to other mathematical and physical concepts.

But conceptual subtleties can also lead to pertinacious misunderstandings. The
most prominent arose from the picture of an "embedding" of this n-component
current theory into its n-component inner symmetry space called
\textit{target} space of string theory. The idea was to convert the
n-component inner symmetry space of the charge-carrying \textit{sigma-model
}$\Psi^{(\vec{q})}(x)$\textit{ fields} into a "target" space which carries a
non-compact group representation, specifically a positive energy
representation of the Poincar\'{e} group.

It turns out that the infinitely many oscillators of a (supersymmetrically
extended) current theory permit precisely one solution in form of the
10-parametric highly reducible \textit{superstring} representation. But in
order to achieve this one has to organize these oscillator degrees of freedom
in a different way from that required by the modular localization on a chiral
lightlike line. In fact already the spectrum of the multicomponent
superselected chiral charge does not match the mass spectrum of the
superstring representation (different zero modes) so that the chiral "source"
theory and the "target representation" of the Poincare group live in different
Hilbert spaces. The localization of a positive energy representation is a fait
accompli and it is point- and not stringlike\footnote{There are no
string-localized infinite spin representation components in the reducible
superstring representation.}. The relation of the chiral model and the
Poincar\'{e} group acting in its putative target space neither support an
embedding of a lower dimensional QFT into a higher dimensional one (which
according to the holistic aspects of different modular localization is
\textit{never possible in QFT}) nor a stringlike localization.

In fact in the sense of loalization in Minkowski spacetime all the non zero
oscillator degrees of freedom form a quantum mechanical oscillator chain in
the inner space "on top of a localization point". One may call this a string
(at the risk of creating confusions), except that in QM localization is not a
spacetime-related intrinsic property but depends on what the working physicist
wants to make out of it. Given the 50 year domination of string theory and the
foundational role of causal localization in QFT these still ongoing
misunderstandings represents certainly the deepest schism which ever occurred
within particle physics.

The source-target terminology incorrectly anticipated that the relation
between the chiral conformal QFT and ST can be understood as an embedding of
the chiral source into the n-component target. Whereas such imbedding (and
their Kaluza-Klein inversions) are perfectly possible in classical field
theories and even in QM, this is not possible in QFT. The reason is that
quantum causal localization is "holistic" whereas localization in classical
field theory or quantum mechanics is not. The holistic organization of
oscillators for implementing the localization of the currents and their
associated sigma model fields on the lightlike line is simply not the same as
that which comes with a positive energy representation and the related
localization on the target.

The only memory which the target space use of the oscillators has about the
chiral current model is that the mass spectrum from the superstring
representation is (up to a scale-setting numerical factor) equal to the
anomalous dimension spectrum of the conformal composites which appear in the
converging global operator expansion of the operator product of two sigma
model fields. This is connected to Mack's observation \cite{Mack1}\cite{Mack2}
that the dual model masses are obtained from the Mellin transform of global
operator product expansions in conformal QFT. Would-be particle spectra in
string theory; the dual model have their origin in anomalous dimension spectra
of conformal QFT and the conformal origin of dual model crossing has no
relation with the crossing in particle physics (section 5) and it is not
comprehensible why the rarity of finding representations of the Poincar\'{e}
group in a target space construction of a chiral theory (the superstring and
its M-theoretic modifications are the only representations) should attach a
fundamental significance which led string theorists to claim that we are
living in a dimensionally reduced 10 dimensional spacetime. The reductionist
trend in QFT supports the idea that a foundational theory should not admit
alternative realizations of the same underlying principle, but it does not
suggest that rarity of a construction should attach a foundational
significance to it.

One can learn a lot from corrections of incorrect ideas about the meaning and
consequences of causal localization in ST. More extensive presentations of
these misunderstandings about localization can be found in \cite{response}.

There are good reasons to also maintain a skeptic attitude with respect to all
ideas which originated from string theory, even if afterwards they were
presented within the setting of QFT as in the case of the AdS-CFT
correspondence. In relations between QFTs in different dimensions the touchy
point is the degrees of freedom issue and its implications for causal
localization; this has simply no counterpart in classical field theory nor in
quasiclassical approaximations. A QFT may fulfill local commutativity
(Einstein causality) but violate the causal completion property as a result of
having more degrees of freedom in the causal completion $\mathcal{A}%
(\mathcal{O}^{\prime\prime})$ than there were in the original region
$\mathcal{A}(\mathcal{O})\varsubsetneq\mathcal{A(O}^{\prime\prime}).$ This
contradicts our ideas of causal propagation; "poltergeist" degree of freedoms
which enter the region of causal determination from nowhere should not occur
in a physically acceptable theory. Their absence is a property of relativistic
propagation of classical Euler-Lagrange equation and enters formally QFT
through Lagrangian quantization; but in a general setting of QFT this has to
be separately added (the time-slice property in \cite{H-S}).

A typical example for a generating field which violates this causality
requirement as the consequence of too many phase space degrees of freedom is
the before mentioned appropriately chosen generalized free field (a free field
with continuous mass distribution). But precisely such fields appears when one
computes the conformal field which according to the AdS-CFT correspondence
results from a AdS free field \cite{Du-Re}. The algebraic derivation of the
AdS-CFT correspondence shows that this phenomenon is intrinsic to this kind of
correspondence. It has its counterpart in the opposite direction
$CFT\rightarrow AdS$, where one finds that there are not enough degrees of
freedom in order to support a physical causal AdS theory, as a result only
non-compact localized algebra as $\mathcal{A}(W)$ are nontrivial whereas
double cone algebras remain empty $\mathcal{A}(\mathcal{C})=\left\{
\mathbb{C}\mathbf{1}\right\}  .$ The preservation of phase space degrees of
freedom is closely related to the shared spacetime group symmetry. This
physical shortcoming in no way has any influence on the mathematical existence
of both sides.\ \ 

These are verifiable structural facts; they do not permit any exception just
like TCP or spin\&statistics. The AdS-CFT Maldacena conjecture, to the extend
that it places a physically viable theory on both sides of the correspondence,
contradicts these facts Even worse, results which once were known at times
when the particle theory community was much smaller seem to have been
irrevocably lost. No wonder that incorrect ideas about embedding of QFTs and
dimensional reduction of extra dimensions enjoy a widespread popularity. In
most publications the awareness that such problems cannot be addressed by
manipulating Lagrangians but need the structural knowledge about interacting
fields and their correlations. Often these misunderstandings arise from
quasiclassical approximations thus overlooking the fact that such
approximations unfortunately do not support modular localization and destroy
quantum degrees of freedom properties. This is especially evident in the
notion of "branes". Their quasiclassical constructions do not show what really
happens namely that \textit{all the degrees of freedom which were contained in
the original physical QFT are compressed into the brane} which, as a result of
overpopulation, becomes unphysical \cite{Mack2}.

One problem in which the relation between localization and cardinality of
phase space degrees of freedom has been used to prove the existence of certain
d=1+1 integrable models is Lechner's work \cite{Lech2}, in particular his
theorem about the nontriviality of double cone intersections of wedges algebra
based on the \textit{modular nuclearity property} of the degrees of freedom
resulting from the Z-F algebra structure of wedge generators.

Acknowledgment: I am indebted to Jens Mund and Jakob Yngvason for innumerous
discussions on various topics which entered this work. Special thanks go to
Detlev Buchholz for a critical reading of section 5 which led to its reformulation.


\begin{thebibliography}{99}                                                                                               %


\bibitem {Arn}V. I. Arnold, V.V. Koslov and A. I.
Neishtadt,\ \textit{Mathematical Aspects of Classical and Celestial
Mechanics}, Springer Verlag 2002

\bibitem {BFK}H. Babujian, A. F\"{o}rster and M. Karowski,
Nucl.Phys.\textbf{B736}, (2006) 169, arXiv:hep-th/0510062

\bibitem {G-J}J. Glimm and A. Jaffe, \textit{Quantum physics: a functional
integral point of view}, Springer Verag 1981

\bibitem {Lech2}G. Lechner, \textit{An Existence Proof for Interacting Quantum
Field Theories with a Factorizing S-Matrix}, Commun. Mat. Phys. \textbf{227},
(2008) 821, arXiv.org/abs/math-ph/0601022

\bibitem {DHN}R.F. Dashen, B. Hasslacher and A. Neveu, Phys. Rev. D.
\textbf{10}, (1974) 4130

\bibitem {STW}B. Schroer, T.T. Truong and P. Weiss, Phys. Lett. B \textbf{63},
(1976) 422

\bibitem {K-W}M. Karowski and P. Weisz, Nucl. Phys. B \textbf{139}, (1978) 445

\bibitem {Borch}H-J Borchers, \textit{On revolutionizing quantum field theory
with Tomita's modular theory}, J. Math. Phys. \textbf{41}, (2000) 860

\bibitem {BKTW}M. Karowski, H.J. Thun, T.T. Truong and P. Weisz, Phys. Lett. B
\textbf{67}, (1977) 321

\bibitem {K}B. Berg, M. Karowski, V. Kurak and P. Weisz, Nucl. Phys. B 134
(1978) 125

\bibitem {Kar2}H. Babujian, A. Fring, M. Karowski, A. Zapletal, Nucl.Phys.
B\textbf{538} (1999) 535

\bibitem {Zam}A. B. Zamolodchikov and Al. B. Zamolodchikov, Ann. Phys.
\textbf{120}, (1979) 253

\bibitem {Ba-Ka}H. Babujian and M. Karowski, Int. J. Mod. Phys. \textbf{A1952}%
, (2004) 34, \ and references therein to the beginnings of the
bootstrap-formfactor program

\bibitem {Sch1}B. Schroer, Nucl. Phys. \textbf{B 499}, (1997) 547

\bibitem {Sch2}B. Schroer, \textit{Modular localization and the d=1+1
formfactor program}, Annals of Physics \textbf{295}, (1999) 190

\bibitem {Lech1}G. Lechner, J.Phys. \textbf{A38}, (2005) 3045

\bibitem {BBS}H. J. Borchers, D. Buchholz and B. Schroer, Commun.Math.Phys.
\textbf{219} (2001) 125

\bibitem {E-J}B. Schroer, \textit{The Einstein-Jordan conundrum and its
relation to ongoing foundational research in local quantum physics}, to be
published in EPJH, arXiv:1101.0569

\bibitem {Du-Ja}A. Duncan and M. Janssen, \textit{Pascual Jordan's resolution
of the conundrum of the wave-particle duality of light}, arXiv:0709.3812

\bibitem {Haag}R. Haag, \textit{Local Quantum Physics}, Springer 1996

\bibitem {Lu-Ma}M. L\"{u}scher and G. Mack, Comm. Math. Phys. \textbf{41},
(1975) 203

\bibitem {Fenn}R. Fenn, R. Rim\'{a}nyi and C. Rourke, \textit{The
braid-permutation group}, Topology \textbf{36}, (1997) 123

\bibitem {Jones}Vaughan F. R. Jones and V. S. Sunder, \textit{Introduction to
subfactors}, Cambridge University Press (1997) Volume 234

\bibitem {Lille}R. Haag, \textit{Discussion of the `axioms' and the asymptotic
properties of a local field theory with composite particles }(historical
document), Eur. Phys. J. H 35, 243--253 (2010)

\bibitem {Wight}R. F. Streater and A. S. Wightman, \textit{PCT
Spin\&Statistics and all that}, New York, Benjamin 1964

\bibitem {Unruh}W. G. Unruh, \textit{Notes on black hole evaporation}, Phys.
Rev. \textbf{D14}, (1976) 870-892

\bibitem {brick}G. 't Hooft, Int. J. Mod. Phys. A\textbf{11,} (1996) 4623

\bibitem {cau}B. Schroer, \textit{Causality and dispersion relations and the
role of the S-matrix in the ongoing research}, arXiv:1107.1374

\bibitem {foun}B. Schroer, \textit{A critical look at 50 years particle theory
from the perspective of the crossing property}, Found. Phys. \textbf{40},
(2010) 1800, \ arXiv:0906.2874

\bibitem {Bi-Wi}J.\ J. Bisognano and E. H. Wichmann, \textit{On the duality
condition for quantum fields}, Journal of Mathematical Physics \textbf{17},
(1976) 303-321

\bibitem {Mund}J. Mund, Commun. Math. Phys. \textbf{286}, (2009) 1159, arXiv:0801.3621

\bibitem {Summers}S. J. Summers, \textit{Tomita-Takesaki Modular Theory},\ arXiv:math-ph/0511034v1

\bibitem {Weinbook}S. Weinberg, \textit{The Quantum Theory of Fields I,
}Cambridge University Press

\bibitem {BGL}R. Brunetti, D. Guido and R. Longo, \textit{Modular localization
and Wigner particles}, Rev. Math. Phys. \textbf{14}, (2002) 759

\bibitem {Fa-Sc}L. Fassarella and B. Schroer, \textit{Wigner particle theory
and local quantum physics}, J. Phys. A \textbf{35}, (2002) 9123-9164

\bibitem {MSY}J. Mund, B. Schroer and J. Yngvason, \textit{String-localized
quantum fields and modular localization}, CMP\textbf{\ 268} (2006) 621, math-ph/0511042

\bibitem {Mu1}J. Mund, J. Math. Phys. \textbf{44} (2003) 2037

\bibitem {nonlocal}B. Schroer, \textit{Unexplored regions in QFT and the
conceptual foundations of the Standard Model}, arXiv:1010.4431

\bibitem {charge}B. Schroer, \textit{An alternative to the gauge theory
setting}, Foun. of Phys. \textbf{41}, (2011) 1543, arXiv:1012.0013

\bibitem {Rio}J. Mund, Prog. Math. \textbf{251}, (2007) 199, arXiv:hep-th/0502014

\bibitem {Mand}S. Mandelstam, Ann. Phys. \textbf{19}, (1962) 1

\bibitem {Scharf}G. Scharf, \textit{Quantum Gauge Theory, A True Ghost story},
2001, John Wiley\& Sons, INC.

\bibitem {Lo-Go}M. Gomes, J. H. Lowenstein, Nucl. Phys. \textbf{B45,} (1972) 252

\bibitem {Sw}A.J. Swieca, Phys. Rev. D \textbf{15}, (1975) 312

\bibitem {Swieca}B. Schroer, \textit{particle physics in the 60s and 70s and
the legacy of contributions \ by J. A. Swieca}, Eur.Phys.J.H \textbf{35},
(2010) 53, arXiv:0712.0371

\bibitem {Higgs}P.W. Higgs, Phys. Rev. Lett. \textbf{12}, (1964) 132

\bibitem {Eng}F. Englert and R. Brout, Phys. Rev. Lett. \textbf{13}, (1964) 321

\bibitem {Gur}G.S. Guralnik, C.R. Hagen and T.W.B. Kibble, Phys. Rev. Lett.
\textbf{13} (1964) 585

\bibitem {L-S}J. H. Lowenstein and B. Schroer, Phys. Rev. \textbf{D7}, (1975) 1929

\bibitem {Due}M. Duetsch, J. M. Gracia-Bondia, F- Scheck, J. C. Varilly,
\textit{Quantum gauge models without classical Higgs mechanism}, arXiv:1001.0932

\bibitem {Ga}J. M. Gracia-Bondia, \textit{On the causal gauge principle}, arXiv:0809.0160

\bibitem {col}The mathematical backup of some of these ideas is part of an
ongoing joint project with Jens Mund and Jakob Yngvason

\bibitem {B-M}J. Bros and J. Mund, \textit{Braid group statistics implies
scattering in three-dimensional local quantum physics}, arXiv:1112.5785

\bibitem {Jakob}J. Yngvason, \textit{The Role of Type III Factors in Quantum
Field Theory}, Rept. Math. Phys. \textbf{55} (2005) 135, arXiv:math-ph/0411058

\bibitem {Jost}R. Jost: \textit{TCP-Invarianz der Streumatrix und
interpolierende Felder}, Helvetica Phys. Acta \textbf{36}, (1963) 77

\bibitem {Ho-Wa}S. Hollands and R. Wald, Gen. Rel. Grav. \textbf{36} (2004) 2595

\bibitem {interface}B. Schroer, \textit{Studies in History and Philosophy of
Modern Physics }\textbf{41} (2010) 104--127, arXiv:0912.2874

\bibitem {Kaehler}R. Kaehler and H.-P. Wiesbrock, \textit{Modular theory and
the reconstruction of four-dimensional quantum field theories}, Journal of
Mathematical Physics \textbf{42}, \ (2001) 74

\bibitem {hol}B. Schroer, \textit{The holistic structure of causal quantum
theory, its implementation in the Einstein-Jordan conundrum and its violation
in more recent particle theories}, arXiv:1107.1374

\bibitem {Sew}G. L. Sewell, Ann. Phys. \textbf{141} (1982) 201

\bibitem {AOP}B. Schroer, Annals Phys. \textbf{275} (1999) 190

\bibitem {Jor}B. Schroer, \textit{Pascual Jordan's legacy and the ongoing
research in quantum field theory,}

\bibitem {1977}D. Buchholz and K. Fredenhagen, Commun. Math. Phys.
\textbf{56}, (1977) 91

\bibitem {DR}S. Doplicher and J. E. Roberts, \textit{Why there is a field
algebra with a compact gauge group describing the superselection structure in
particle physics}, Commun. Math. Phys. \textbf{131}, (1990) 51-107

\bibitem {Re-Sch}K-H. Rehren and B. Schroer, Phys. Lett. B\textbf{198,
(}1987\textbf{) 84}

\bibitem {old}B. Schroer, \textit{Space- and timelike superselection rules in
conformal quantum field theories}, hep-th/0010290 (2000), see also B. Schroer,
\textit{Braided structure in 4-dimensional quantum field theory}, Phys. Lett.
B, \textbf{506,} (2001) 337

\bibitem {S-S}B. Schroer and J. A. Swieca, Phys. Rev. D \textbf{10}, (1974) 480

\bibitem {S-S-V}B. Schroer, J. A. Swieca and A. H. Voelkel, Phys. Rev. D
\textbf{11}, 1975

\bibitem {Mund3}J. Mund, J. Math. Phys. \textbf{44} (2003) 2037

\bibitem {Ni-Re-To}N. M. Nikolov, K.-H. Rehren and I. Todorov, Commun. Math.
Phys.\textbf{279}, (2008) 225

\bibitem {BPZ}A. A. Belavin, A. M. Polyakov and A. B. Zamolodchikov, Nucl.
Phys. B \textbf{241} (2), (1984) 333

\bibitem {Bei}N. Beisert, Nucl. Phys. B 682 487 (2004) arXiv:hep-th/0310252

\bibitem {R-S}K.-H. Rehren and B. Schroer, \textit{Einstein causality and
Artin braids}, Nucl. Phys. B \textbf{312}, (1989) 715

\bibitem {F-G}J. Fr\"{o}hlich and F. Gabbiani, Braid statistics in local
quantum theory, Rev. Math. Phys. \textbf{2}, (1991) 251

\bibitem {Mack1}G. Mack, \textit{D-dimensional Conformal Field Theories with
anomalous dimensions as Dual Resonance Models}, arXiv:0909.1024

\bibitem {Mack2}G. Mack, \textit{D-independent representation of Conformal
Field Theories in D dimensions via transformation to auxiliary Dual Resonance
Models. Scalar amplitudes}, arXiv:0907.2407v1

\bibitem {Ka-Lo}Y. Kawahigashi and R. Longo, Commun. Math. Phys.\textbf{244}
(2004) 63

\bibitem {response}B. Schroer, \textit{The holistic structure of causal
quantum theory, its implementation in the Einstein-Jordan conundrum and its
violation in more recent particle theories}, arXiv:1107.1374

\bibitem {H-S}R. Haag and B. Schroer, \textit{Postulates of Quantum Field
Theory}, J. Mat. Phys. \textbf{3}, (1962) 248

\bibitem {Du-Re}M. D\"{u}tsch and K.-H. Rehren, Ann. Henri Poincare
\textbf{4}, (2003) 613
\end{thebibliography}
\end{document}